\documentclass{aastex62}

\usepackage{amsmath,amssymb}
\usepackage{float}
\usepackage{color}

\def\deg{\ifmmode^\circ\else$^\circ$\fi}

\def\refcat{Refcat2}

\graphicspath{{./}{figures/}}

\received{January 1, 2018}
\revised{January 7, 2018}
\accepted{\today}
\submitjournal{ApJ}

\shorttitle{ATLAS Reference Catalog}
\shortauthors{Tonry et al.}

\begin{document}

\title{The ATLAS All-Sky Stellar Reference Catalog}

\correspondingauthor{John Tonry}
\email{tonry@hawaii.edu}

% Authors
\author{J. L. Tonry}
\affiliation{Institute for Astronomy, University of Hawaii, 2680 Woodlawn Drive, Honolulu, HI 96822}
%\email{tonry@hawaii.edu}

\author{L. Denneau}
\affiliation{Institute for Astronomy, University of Hawaii, 2680 Woodlawn Drive, Honolulu, HI 96822}
% \email{denneau@hawaii.edu}

\author{H. Flewelling}
\affiliation{Institute for Astronomy, University of Hawaii, 2680 Woodlawn Drive, Honolulu, HI 96822}
% \email{heather@ifa.hawaii.edu}

\author{A. N. Heinze}
\affiliation{Institute for Astronomy, University of Hawaii, 2680 Woodlawn Drive, Honolulu, HI 96822}
% \email{aheinze@hawaii.edu}

\author{C. A. Onken}
\affiliation{Research School of Astronomy and Astrophysics, Australian National University, Canberra, ACT 2611, Australia}
% \email{christopher.onken@anu.edu.au}

\author{S.J. Smartt}
\affiliation{Astrophysics Research Centre, School of Mathematics and Physics, Queen's University Belfast, Belfast BT7 1NN, UK}
% \email{s.smartt@qub.ac.uk}

\author{B. Stalder}
\affiliation{LSST, 950 N. Cherry Ave, Tucson, AZ 85719}
% \email{bstalder@lsst.org}

\author{H. J. Weiland}
\affiliation{Institute for Astronomy, University of Hawaii, 2680 Woodlawn Drive, Honolulu, HI 96822}
% \email{hweiland@hawaii.edu}

\author{C. Wolf}
\affiliation{Research School of Astronomy and Astrophysics, Australian National University, Canberra, ACT 2611, Australia}
% \email{christian.wolf@anu.edu.au}

\begin{abstract}

The Asteroid Terrestrial-impact Last Alert System (ATLAS) observes
most of the sky every night in search of dangerous asteroids.  Its data
are also used to search for photometric variability, where sensitivity
to variability is limited by photometric accuracy.  Since each exposure spans
7.6\deg\ corner to corner, variations in atmospheric transparency
in excess of 0.01 mag are common, and 0.01 mag photometry cannot be achieved
by using a constant flat field calibration image.  We therefore have
assembled an all-sky reference catalog of approximately one billion stars 
to $m\sim 19$ from a variety of
sources to calibrate each exposure's astrometry and photometry.
Gaia DR2 is the source of astrometry for this ATLAS
\refcat.  The sources of $g$, $r$, $i$, $z$
photometry include Pan-STARRS DR1, the ATLAS Pathfinder
photometry project, ATLAS re-flattened APASS data, SkyMapper DR1, APASS DR9, the
Tycho-2 catalog, and the Yale Bright Star Catalog.  We have attempted to make
this  catalog at least 99\% complete to $m<19$, including the brightest stars in the sky.
We believe that the systematic errors are
no larger than 5~millimag RMS, although errors are as large as 
20~millimag in small patches near the galactic plane.
\end{abstract}


\section{INTRODUCTION}
\label{sec:intro}

The Asteroid Terrestrial-impact Last Alert System
has been funded by NASA to find dangerous asteroids
that might threaten the Earth
\citep{2018PASP..130f4505T}. 
Henceforth known as the ATLAS project, it requires extremely accurate
photometry and astrometry for a number of reasons.
\begin{itemize}
  \item{} Good astrometry is critical for determining precise minor planet orbits.
    ATLAS achieves a positional accuracy of $\sim 0.07\arcsec$ RMS
    per star relative to a fit to a frame or a astrometric reference for $m<17$, 
    so without an absolute astrometric reference catalog (such as Gaia) that is 
    substantially better than that we cannot provide the Minor Planet 
    Center (MPC) with measurements of asteroid positions that exploit the
    full potential of our system for precise astrometry. Such homogeneous,
    precise measurements will ultimately aid the detection and measurement of
    non-gravitational forces on asteroid orbits, e.g. the Yarkovsky
    effect.
  \item{} From good photometry we can derive extremely accurate
    light curves, periods, and colors for asteroids 
    \citep[see][as an example from Pan-STARRS data]{2015Icar..261...34V}.     
    In the long run
    this will allow non-gravitational torques to be measured
    \cite[due, e.g., to the YORP effect;][]{2007Sci...316..272L}
  \item{} We detect asteroids by subtracting a static ``wallpaper''
    image of the sky, and that subtraction does not reach photon
    limited performance unless the astrometry is precise to 
    at least 0.05 pixels and the photometry to at least 0.03 mag.
  \item{} As a byproduct of the asteroid search ATLAS also produces
    light curves for every other detected object in the sky, whose value
    for a myriad of scientific studies depends on
    the quality of the photometry.  There is rich discovery space for
    time-domain studies of stars, galaxies and transients: ATLAS
    data are already contributing through the first release of a 4.7
    million variable star catalog.  \citep{2018arXiv180402132H}.
  \item{} A particular challenge for calibrating wide-field images is
    patchy, very thin clouds, common across the 30
    square degree field-of-view in ATLAS. We address this
    challenge by identifying $\sim 10^5$ stars in each frame and
    deriving a ``cloud correction'' by comparing them to an accurate
    reference catalog. Such patchy obscuration has been detected
    before when hyper-calibration of large, well calibrated data has
    been achieved \citep[e.g. the contrail effect visible when
      Pan-STARRS and the Sloan Digital Sky Survey data were combined
      by][]{2016ApJ...822...66F}.
\end{itemize}

An all-sky photometric catalog with the accuracy
required by ATLAS is not currently available. While the ESA/Gaia
mission \citep[Data Release 2, DR2, overview description][]{2018arXiv180409365G} is
expected to provide the groundwork for such a catalog at an
unprecedented level of precision, it will not support ATLAS bandpasses
until the release of low-resolution spectra expected in their DR3 by
the year 2021.  Since ATLAS needs a precise reference catalog now, the
purpose of this work is to construct it from a variety of existing
sources, and we expect this catalog to be beneficial to other projects as
well.  This paper is part of a
series that describe the ATLAS project and its various components and
data products, the first two being the system definition paper
\citep{2018PASP..130f4505T} and the first variable star data release
\citep{2018arXiv180402132H}.

There are many sources of optical photometry, none of which provides
the perfect reference source for ATLAS observations, which currently
range from the north celestial pole to Dec $-50$\deg, include
the Galactic plane and extend as bright as $m\sim0$.  ATLAS has 
also started on two additional units in South Africa and Chile.
Contemporary surveys most often employ $griz$-filter systems, 
similar to bandpasses used in the Sloan Digital Sky Survey
\citep[SDSS;][]{Fukugita96, Fukugita11} and Pan-STARRS
\citep{psphot,2016arXiv161205560C,2016arXiv161205243F}. Hence, it seems
natural to construct a reference catalog with $griz$ photometry and
then apply mild color transformations when using it to calibrate
observations using different filter sets. We initially started
calibrating ATLAS data with Gaia Data Release 1
\citep[DR1;][]{2016A&A...595A...2G}, Pan-STARRS Data Release 1
\citep[DR1;][]{2016arXiv161205242M,2016arXiv161205560C}, and APASS Data Release 9
\citep[DR9;][]{2016yCat.2336....0H}, but a number of issues cause
problems.  Gaia DR1 has holes in its coverage, Pan-STARRS DR1 is not
accurate brighter than $g<14$ and does not extend south of Dec
$-30\deg$, and APASS DR9 photometry is not reliable at the 0.05 mag
level. To exploit ATLAS data optimally, we therefore require a new
all-sky reference catalog that meets our own strict requirements.

We spent a year collecting $gri$ images with a small ATLAS Pathfinder
Telescope on Mauna Loa in order to obtain photometry for stars
brighter than the $m\sim14$ magnitude limit of Pan-STARRS and to push our 
photometric reference south of Dec $-30$\deg.
In addition, the AAVSO/APASS team very generously
sent us a nearly complete set of images covering the sky
from the south pole to Dec $+20\deg$ which we re-flattened,
re-photometered, and re-combined into a new southern sky catalog. 

Gaia DR2 is a truly beautiful product.  The
completeness is excellent; few stars brighter than $m<18$ are
missing.  The astrometry (and proper motions and parallaxes) are
superb, of course.  In addition, Gaia DR2 offers $G_{BP}$ and $G_{RP}$
photometry which appears to be extremely uniform and accurate.
Another important, recent data source is the SkyMapper Data Release 1
\citep[DR1,][]{Wolf18} which provides $griz$ for most of the southern
sky.  Although its photometry is currently tied to APASS DR9, it goes
considerably deeper and is more accurate. Of course, 2MASS
\citep{2MASS} is a well calibrated all-sky resource as are the Tycho-2
catalog \citep{2000A&A...357..367H} and the Yale Bright Star Catalog
\citep{1964cbs..book.....H}.

In this paper we present a compilation of these sources to produce an
all-sky catalog of stellar photometry that we believe is accurate
and virtually complete to $m\sim19$.  The catalog data are based on our
new observations, our re-reductions, and the extant catalog data,
synthesizing all the $griz$ photometry into a best
estimate of $griz$ for each star on the Pan-STARRS bandpasses.

We will describe our
Pathfinder observations and reductions, our APASS re-reductions, the
procedure we followed for each catalog to bring it onto the
Pan-STARRS $griz$-bandpass system, and the production of the final
catalog.  We call this final data product ATLAS \refcat.  The precise
content of \refcat\ is described in Appendix~\ref{app:details}.

\section{External sources of photometry}

\subsection{2MASS}

The 2 Micron All Sky Survey \citep[2MASS,][]{2MASS} covered the entire
sky in infrared $JHK_{\rm s}$ bandpasses.  The systematic errors in astrometry
are no larger than 0.05\arcsec\ and in photometry no larger than 0.02 magnitude, but the
infrared magnitudes are not directly convertible to $gri$ at the level
required here.  These magnitudes are very useful in combination with
the Gaia photometry to distinguish stellar color from reddening,
however.  We downloaded the 2MASS data from IPAC{\footnote{\tt
    http://irsa.ipac.caltech.edu/2MASS/download/allsky}}.

\subsection{APASS Data Release 9}

The AAVSO has released a catalog of stellar photometry for about 50
million stars over the entire sky, from the AAVSO Photometric All Sky
Survey \cite[APASS DR9;][]{2016yCat.2336....0H}.  The astrometry is
tied to UCAC3 \citep{2010AJ....139.2184Z}, with an accuracy better than
0.1\arcsec, and its photometry includes $BVgri$ across the range
$7<m<17$.  Their cameras are Apogee U16m with a 4k KAF16803 detector
and Astrodon $u^\prime$, $g^\prime$, $r^\prime$, $i^\prime$,
$z^\prime$, $B$, and $V$-filters.  The KAF16803 detector has
lower quantum efficiency
than our ATLAS Pathfinder CCID20 CCDs (see Section\,\ref{sec:pt}), 
but APASS used 180/90/90~sec
exposures in $gri$, so the depths are comparable.

As described below, we corrected for substantial, systematic 
photometric offsets in this published DR9 dataset.  Because 
APASS carried out a flattening correction, photometry, and detection
combination for DR9 that is independent of the work performed here, we
consider it to be a quasi-independent source of photometry and we added it
into the sources of $gri$ magnitudes, once these offsets were
corrected on a square degree basis.

\subsection{Pan-STARRS}
\label{sec:ps1}

The Pan-STARRS1 Telescope \citep{2016arXiv161205560C} surveyed the sky
north of Dec $-30$\deg\ many times in $grizy$ filters between 2010 and 2014.
The Pan-STARRS bandpasses have been well characterized \citep{psphot}.
These data have been homogenized using an ``ubercal'' procedure
\citep{Padmanabhan08,Schlafly12} and tied to the Space Telescope
Science Institute's ``Calspec''  stars
\citep{Calspec,Bohlin07,Bohlin14}. 
The Pan-STARRS1 data processing \citep{2016arXiv161205240M,2016arXiv161205245W,2016arXiv161205244M}
and  calibration \citep{2016arXiv161205242M}
provide accurate $grizy$ photometry ($<0.005$ mag) and astrometry ($<20$ milliarcsec),
but only for stars fainter than $m\sim14$; brighter stars are saturated.

We used the Pan-STARRS DR1 catalog database
\citep{2016arXiv161205243F}, by running a query that checks that the
object has a stack magnitude less than 19 in any of $g$,$r$,$i$, or
$z$, and that {\tt bestDetection=1} and {\tt primaryDetection=1} in
the {\tt StackObjectThin} tables, and reports back the mean positions
(RA/Dec) and mean PSF (point spread function) magnitudes in $grizy$ from the {\tt ObjectThin}
and {\tt MeanObject} tables.  This query was chosen to restrict the
objects to those that are bright enough for ATLAS as well as to select
objects that have been seen multiple times by Pan-STARRS and for which
the Pan-STARRS stack photometry (measurement from the co-add of all images) succeeded.  
We use the Pan-STARRS mean
photometry (mean of measurements from all images) because it is 
has had ubercal corrections applied \citep[see][]{Schlafly12,2016arXiv161205242M}
whereas the corresponding stack
photometry has not.  However to maximize the probability that an object
is a real star (and not a spurious detection, transient source or a
moving object), it is useful to check that it exists and that
certain flags are set in the stack table. In Appendix\,\ref{app:ps1} we provide an example of the
query, should others wish to select a similar set of Pan-STARRS1 data
from the MAST archive\footnote{\tt http://mastweb.stsci.edu/ps1casjobs/}. 

\subsection{SkyMapper}

In Dec 2017, SkyMapper published a slight revision to their first data
release, DR1.1\footnote{\tt http://skymapper.anu.edu.au/data-release},
comprising 285 million sources \citep{Wolf18}.  DR1.1 employs 2MASS
and a selection of APASS stars to set observation zeropoints
(the number that converts flux reported by photometry routines to
calibrated magnitude) and determine magnitudes.  We found that
photometric errors in APASS DR9 had left a substantial imprint on
SkyMapper photometry, so we corrected the SkyMapper photometry on each
square degree, as described below. We caution that the point-source
photometry in DR1.1 is derived from 1D-growth curves and thus is only
reliable for isolated sources that have no neighbors within
$10\arcsec$.

\subsection{Tycho-2 and bright stars}

The ESA Hipparcos satellite \citep{2007A&A...474..653V} 
surveyed the entire sky to determine parallaxes of bright stars, and it included the
Tycho photometry instrument.  The Tycho-2 catalog it produced
contains almost all the stars in the sky brighter than $m\sim12$
\citep{2000A&A...357..367H}.  Its astrometry is superb and its
photometry is quite homogeneous over the entire sky, albeit
subject to crowding problems in the galactic plane.  Tycho had two
filters $B_T$ and $V_T$, similar to Johnson $B$ and $V$, so cannot be
compared directly with $gri$ photometry without significant color
transformations.

The 2MASS team performed a match between Tycho-2 and 2MASS, 
providing $B_TV_TJHK$ photometry for about 2.5 million stars.
\citet{Pickles2010} then used this 2MASS match to determine a best-fitting
stellar spectral energy distribution (SED) from their compendium of SEDs, 
and published synthetic $gri$ photometry for each of the Tycho-2 stars.  

The Tycho-2 catalog becomes quite incomplete for $m<3$, so we
augmented it with the Yale Bright Star Catalog (BSC)
\citep{1964cbs..book.....H,1995yCat.5050....0H}.
%SJS - as CW notes, there does not seem to be a relevant 1991 citation here. 
%JT to add or leave deleted. 
%{\bf REF Hoffleit and Warren 1991, [CW: what is 1991 ?]}.  
$griz$ magnitudes were estimated from the $B$, $V$,
and $(U-B)$ and $(R-I)$ colors when available.  These estimates
are {\it not} particularly accurate, but they ensure the completeness of
the final \refcat.
We shifted Tycho-2 and BSC coordinates to
epoch 2015.5 according to the Tycho-2 proper motions and merged the
two, with preference to Tycho-2 for stars listed by both.

\subsection{Gaia}

The Gaia Collaboration published a first data release in 2016,
\citep[DR1][]{2016A&A...595A...2G} and issued a greatly improved second data
release DR2 on 25 Apr 2018 \citep{2018arXiv180409365G}.  DR2 has three substantial
improvements relative
to DR1: it closes holes in the sky, it provides Gaia ``$G_{BP}$''
and ``$G_{RP}$'' magnitudes for stars brighter than $G\sim19$,
and it provides proper motions and parallaxes for many stars. The star
coordinates in DR2 are all epoch 2015.5.  This is a revolutionary
dataset, not only because of the unprecedented depth and accuracy of
the astrometry, but also because of the accuracy of the photometry.
Comparison of DR1 $G$ magnitudes against Pan-STARRS clearly showed the
Gaia scan patterns, indicating photometry errors at the $\sim0.02$
magnitude level.  DR2 does not show Gaia scan patterns relative to
Pan-STARRS at the $0.005$ magnitude level or better, and of course
there is every expectation that the Gaia photometry can be
homogeneous over the entire sky. As we discuss later (Section\,\ref{sec:GMP}), we 
do observe
a small difference between Gaia DR2 photometry and ground-based photometry
that correlates with star density.

\section{New sources of photometry data and our processing}

In addition to extant catalogs, our reference catalog also includes a
new photometric information over much of the sky from our own ATLAS Pathfinder 
observations as well as from a re-processing of APASS images.

\subsection{ATLAS Pathfinder Hardware and Survey}
\label{sec:pt}
The ATLAS Pathfinder Telescope is a Takahashi Epsilon-180
astrograph that has a 5 degree diameter field of view, an aperture of 180mm and
a focal length of 500mm.  It is equipped with an FLI PDF focuser and
an FLI filter changer.  The filter changer has
Astrodon\footnote{\tt http://www.astrodon.com/sloan.html} 50mm SDSS Gen-2
$g$, $r$, and $i$-band filters inserted.  These filters have very square
bandpasses with half-transmission points of 401--550nm ($g$-filter),
562--695nm ($r$-filter), and 695--844nm ($i$-filter).  This is followed by
a Uniblitz 45mm shutter and a custom CCD camera.
The CCD camera has two 2k$\times$4k Lincoln Lab CCID20 CCDs with
15~um pixels, for a plate scale of 6.2 arcsec/pixel.  The Takahashi
illuminates a circle of diameter 3000 pixels, so the field of
view is about 19 square degrees.  These CCDs are
back-illuminated and 45~um thick so their quantum efficiency and
uniformity is excellent.  

The telescope was used on Mauna Loa 
in the ATLAS Ash dome at the NOAA observatory at an
elevation 11,000 feet for approximately 2 years, prior to the
installation of the ATLAS 0.5~m telescope. 
We observed between MJD 57396 (2016-01-09) and 57782 (2017-01-29),
initially collecting images in a single filter on a given night,
applying dithers and letting some time elapse between observations.
In March 2016 (MJD 57449) we switched to a mode of collecting three consecutive images
in $gri$ at each pointing, covering about 1/5 of the visible
sky on a given night.  We normally observed in five Declination bands
between $-45$ and $+90$. All exposure times were 30~sec, and the Takahashi 
vignettes about 40\% of the light at the edge of the field of view.

The Takahashi telescope produces a focal surface that curls up at the
edge of the field of view, and at f/2.8 there is a distinct
non-uniformity of focus.  In addition the detectors are slightly
tilted with respect to the focal surface.  At its best the Takahashi
makes images of about 1.2 pixels, but we deliberately chose a
focus that produces images of about 1.7 pixels full width at half maximum (FWHM) over the entire
field of view.  There is a distinct variation in the shape of the
images as a function of position, but the overall FWHM is reasonably
constant for all filters over the entire field of view.  However,
prior to running {\tt DoPhot} we elected to convolve images with a Gaussian 
of 2 pixels FWHM.  This costs about a half magnitude in sensitivity but
significantly improves the photometric accuracy over the field.

\begin{figure}[htbp]
\begin{center}$
\begin{array}{ccc}
\includegraphics[width=3in]{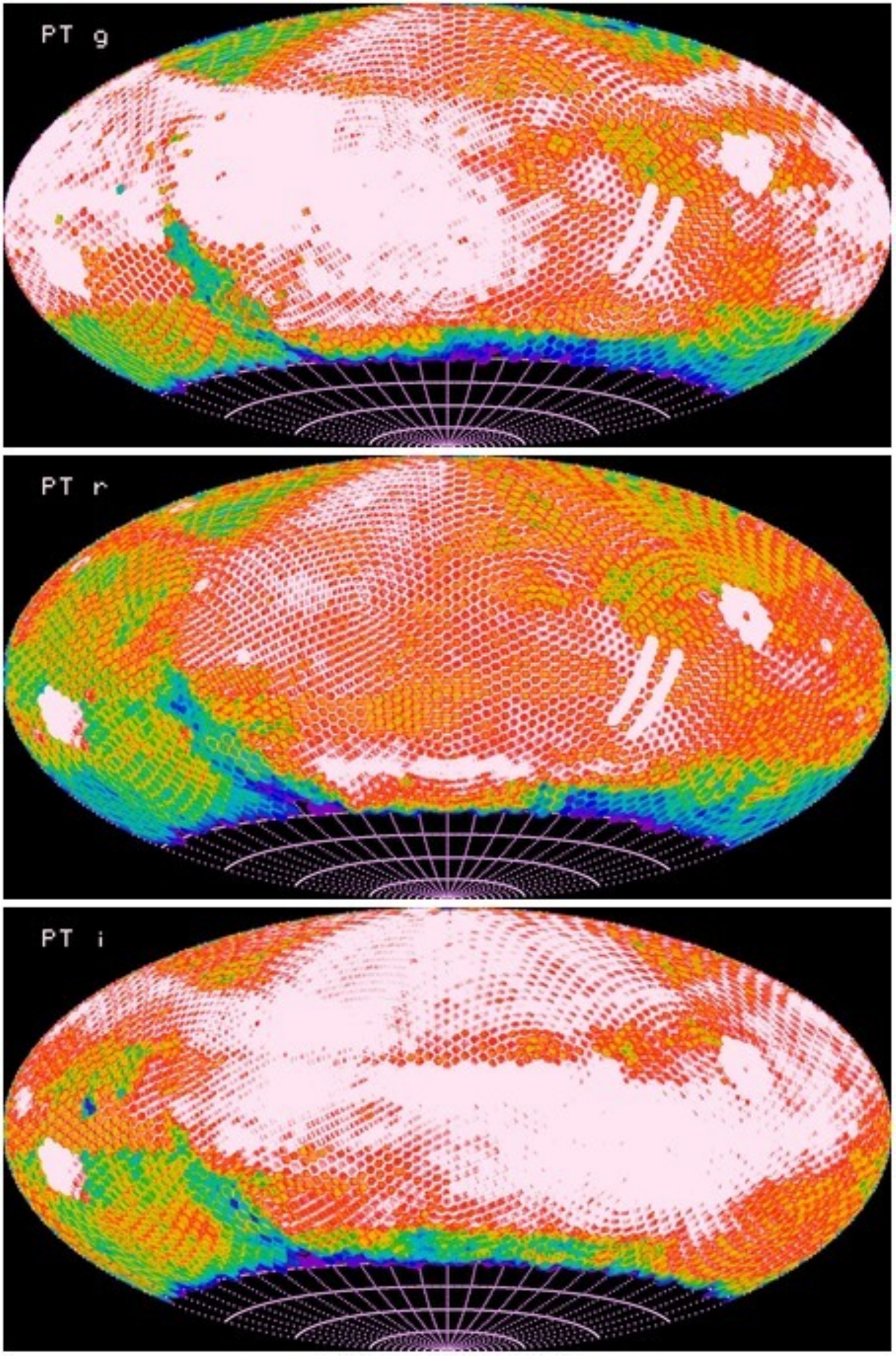}
\qquad
\includegraphics[width=3in]{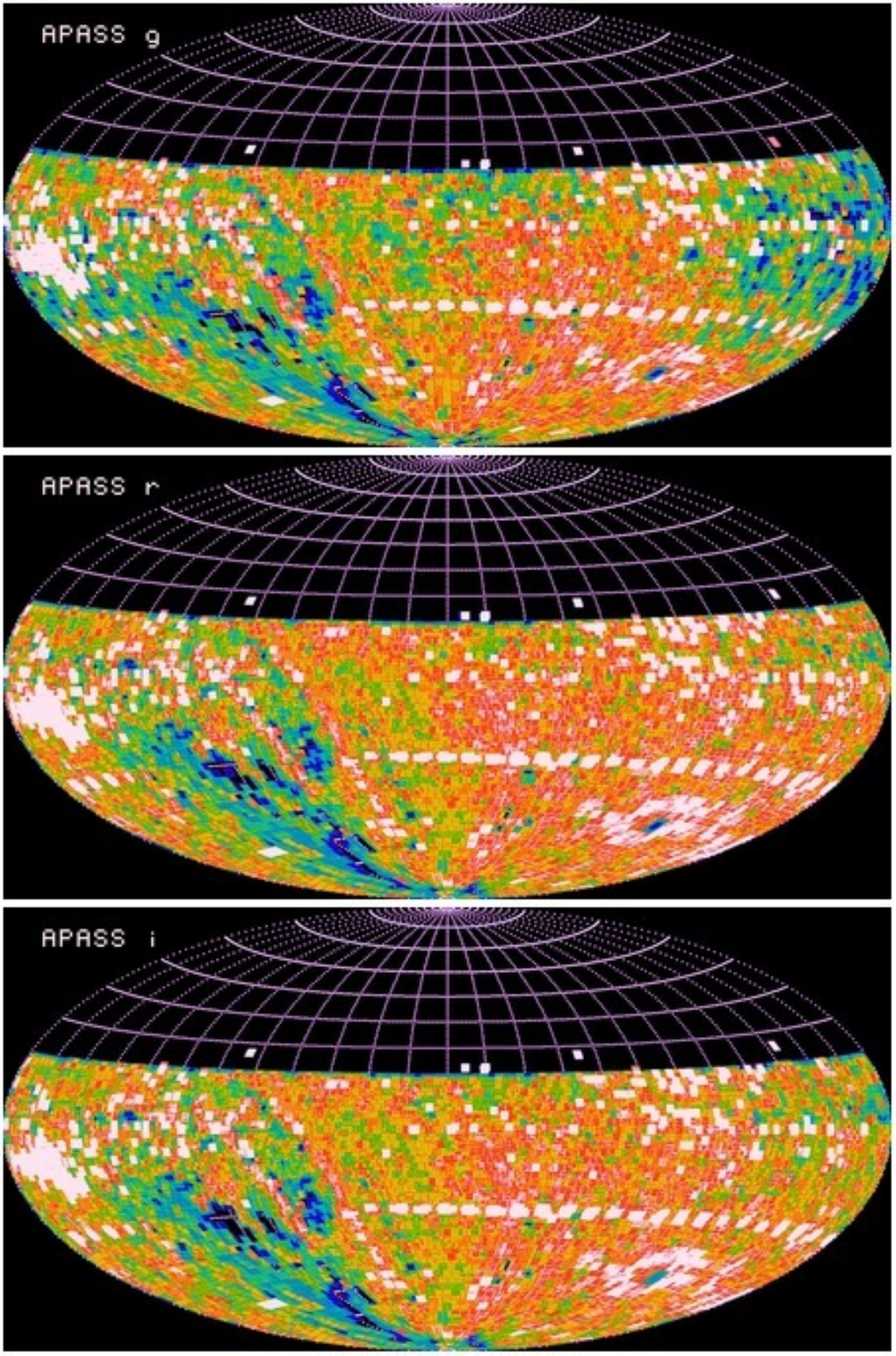}
\end{array}$
\end{center}
\caption{The coverage of the sky by ``good'' Pathfinder exposures
  is illustrated on the left with the number of visits coded by color:
  blue is 5, yellow is 15, red is 20, and white is
  28 or more.  Filters $g$, $r$, and $i$ run from top to bottom,
  RA increases to the left.  The right panel shows the sky coverage by 
  ``good'' APASS observations, the number of
  visits coded by color: blue is 2, yellow is 5, red is 8, and white is
  9 or more.}
\label{fig:ptcover}
\end{figure}

Among the Pathfinder data, ``good nights'' were identified as those
with a scatter in zeropoint (magnitude that provides 1 count per second)
less than 0.03~mag (4/3 the quartile range)
as a function of airmass over the majority of the night, and ``good
observations'' were selected.   A total of 362 nights altogether
and 232,558 exposures in $gri$ yielded 213 nights that were at least
partly photometric, with a total of 165,294 exposures.  
After fairly stringent cuts on quality control
metrics 159,397 exposures survived to provide stellar photometry.
Our sky coverage of these exposures is displayed in Figure~\ref{fig:ptcover}.
Each image produces a table of about 30,000
stars with astrometric accuracy of about 0.2\arcsec\ relative
to Gaia DR1 and photometric accuracy of about 0.05 mag relative to Pan-STARRS
at $g\sim16.5$, $r\sim16$, and $i\sim15.5$.  
The Pathfinder observations saturate at approximately $m<9$.

\subsection{Photometric processing of ATLAS Pathfinder data}

The basic ATLAS pipeline \citep{2018PASP..130f4505T}
was used to reduce the ATLAS 
Pathfinder data.  The stages
involve bias subtraction, division by a flatfield, and identifying many
stars 
to determine pixel positions and fluxes.
A subset of these stars are matched to sky positions using code from {\tt
  astrometry.net} \citep{2010AJ....139.1782L} to obtain an initial
astrometric solution. All stars are then matched to the first generation ATLAS
reference catalog\footnote{ATLAS project internal version Refcat1 based on Gaia DR1, Pan-STARRS, and APASS DR9} to derive an accurate astrometric solution expressed as WCS
coefficients and a photometric zeropoint. 
Refcat1 has estimates of star magnitudes on the  Pan-STARRS $g_{P1}$, $r_{P1}$,
and $i_{P1}$ filter system \citep{psphot},
and these magnitudes are  
converted to the Pathfinder $gri$ observational bandpasses (see Table~\ref{tab:cterm} below).

The flatfield is initially a night time sky flat image, derived from the median of
all images in a given filter for a night, but then the stars from
images that overlap the Pan-STARRS catalog are compared with
Pan-STARRS magnitudes, a low order $8\times8$ spline fit to the
differences is multiplied into the flatfield to make a ``photo-flat''
or ``star flat'', and the data from the night are re-reduced with this
new flatfield.
As described below, the purpose of the photo-flat is to correct
intra-exposure photometric variations, {\it not} to get the final
zeropoint for the exposure correct; the zeropoint evaluation follows a
completely different procedure.  This photo-flat serves for a lunation.

Finally we run a version of {\tt DoPhot} \citep{DoPhot} that
was modified by Alonso-Garcia \citep{dophotf90} to accept
floating point data and to cope with a varying point spread function.
We also had to make a number of modifications to {\tt DoPhot} as well to get
the variable PSF to work properly and also to be able to introduce an
external variance image.
{\tt DoPhot} processes an image iteratively.  At any point it maintains a
catalog of all stars it has detected, and it subtracts its model of
those stars from the image before detecting and measuring new stars or
remeasuring old stars in the catalog.  In this way {\tt DoPhot} does a very
creditable job of dealing with high star densities and blended images.

{\tt DoPhot} calculates two different fluxes for most stars: an ``aperture
magnitude'' which is the sum of the flux within a large aperture
($\sim30$\arcsec) radius, and a ``fit magnitude'' which is
the flux derived from the integral of the {\tt DoPhot} PSF function.  The
former includes almost all the light from a star but is noisy, the
latter has systematic errors when the {\tt DoPhot} PSF model does not match
the actual star profile.  All stars have a ``fit magnitude'', but only
the bright stars have an ``aperture magnitude''.  By examining the
difference between aperture and fit magnitudes for the bright stars,
the fit magnitudes for faint stars can be converted to aperture
magnitudes.  The result is that we can produce low noise instrumental
magnitudes for each star that are referred to large aperture fluxes.

\subsection{APASS re-reduction}

The APASS project kindly provided ATLAS with 258,312 images of the
southern sky on a 3TB external drive.  They were taken at the APASS
facility at the Cerro Tololo Inter-American Observatory in Chile,
primarily (237,498 images) in $g$, $r$, $i$, $B$, and $V$ filters over
a span of MJD 55507 (2010-11-07) to 56725 (2014-03-09).  They cover
nearly the entire sky with Dec $<+20$\deg, with few gaps (the
southern ``crack in the sky" that stretches between $\alpha,\delta$
$260\deg,-60\deg$ to $270\deg,-77\deg$ is not covered by these images). 
The APASS system has a 2.9$\times$2.9\deg\ field of view with
2.6\arcsec\ pixels, and exposure times were varied so as to extend the
dynamic range to brighter stars.

We reduced all APASS exposures through the ATLAS pipeline, including
the match with  ATLAS reference catalog Refcat1.
Because the APASS observing
strategy on each night brought the telescope well north of Dec
$-30$\deg, virtually every night overlapped the Pan-STARRS catalog
well enough to provide a photo-flat correction to the APASS
flatfield.
This was not a trivial modification, amounting to a
peak-to-peak of 0.2 magnitude over a typical image and an RMS of
about 0.05 mag.  There was also a dramatic, steady decrease in zeropoint
by 0.7 mag in all filters over this 3.3 year span.  We found that the
Pan-STARRS flatfield comparison was very consistent in a given filter
from night to night, and on the few nights when APASS did not venture
north of $-30$\deg\ we used the photo-flat correction from the closest
night that did.
The astrometric errors are typically 0.07\arcsec\ in each coordinate
at $m\sim15$, and the typical photometric errors in the long exposures
are 0.05 mag at $g\sim17.5$, $r\sim16.5$, $i\sim15$, $B\sim17.5$, and
$V\sim16.5$. 

As with the Pathfinder data, ``good nights'' and ``good observations''
were identified from the scatter in zeropoint as a function of airmass,
and quality control metrics.  692 nights altogether yielded 542 nights
that we deem were photometric, with a total of 178,094 exposures of which
154,406 in $g$, $r$, $i$, $B$, and $V$ provided stellar photometry.
The APASS sky coverage is displayed in the right panel of Figure~\ref{fig:ptcover}.
(There is also an APASS-north survey which we did not use.)

{\it It is important that readers and users note the following.}  
Hereafter the ``APASS'' catalog will be used to
mean the APASS data that has been reprocessed by ATLAS, and the
``APASS DR9'' catalog will refer to the magnitudes published by the
APASS project.  These share many of the same data files but the
processing is completely different.  As detailed below, this
re-reduction of APASS made the intra-CCD photometry about a factor of
3 better than what was published in APASS DR9, the re-processed
photometry is about 2 magnitudes deeper, and even after we correct
APASS DR9 into agreement with Gaia and Pan-STARRS this re-reduced
APASS catalog is distinctly more homogeneous across the sky.

\section{Zeropoints}

Table~\ref{tab:zp} shows catalog comparison with Pan-STARRS in the
same equatorial band between $-28$\deg \ and $+18$\deg \ Dec, collecting
median differences of star magnitudes over each coordinate square degree and then
examining the statistics between these square degrees.  Color terms
from Table~\ref{tab:cterm} have been applied to each catalog's magnitudes 
to bring them into agreement with Pan-STARRS.

Note that APASS DR9 has substantial offsets with respect to Pan-STARRS
(53 millimag in $g$ for example), and substantial RMS values of
11--17~millimag.  Similarly, SkyMapper DR1.1 is better corrected in terms
of zeropoint and was distinctly tighter (RMS of 6--10~millimag), but the
large scale differences between APASS and Pan-STARRS are echoed in
SkyMapper DR1.

\begin{table}[htp]
\caption{Survey zeropoints}
\begin{center}
\begin{tabular}{lcrrrrrrrr}
\hline
Survey          & ZP   & $g$&$dg$& $r$ &$dr$& $i$ &$di$& $z$ & $dz$\\
\hline                    
APASS DR9       & ---  & 53 & 11 & -13 & 12 & -34 & 17 &     &    \\
SkyMapper DR1.1 & ---  & -5 & 10 & -18 & 10 &  -9 &  6 &  10 &  8 \\
APASS           & diff &  0 & 17 &   0 & 10 &   0 & 12 &     & \\
APASS           & secz &  0 & 13 &   0 &  7 &   0 & 11 &     & \\
APASS           & GMP  &  0 &  2 &   0 &  1 &  -1 &  2 &     &    \\
Pathfinder      & GMP  & -1 &  1 &  -0 &  1 &  -1 &  2 &     &    \\
SkyMapper       & GMP  & -2 &  2 &  -3 &  2 &  -2 &  2 &   0 &  1 \\
APASS DR9       & GMP  & -3 &  2 &  -2 &  2 &   1 &  2 &     &    \\
\hline
\end{tabular}
\end{center}
\tablecomments{For each survey the median difference in millimag with
  respect to Pan-STARRS for $-28<\delta<+18$ is given in $gri$ filters
  and $z$ if available.  These differences are calculated from the
  median of matched stars over each coordinate square degree, then the
  median of these and their RMS (derived from quantiles) is listed.
  The left column gives the survey and the next column lists the
  source of zeropoint recalibration, discussed in this section.}
\label{tab:zp}
\end{table}%

We regard an RMS of 10~millimag as unacceptable for our reference
catalog because it is a {\it systematic} error that would be impressed
on every observation that used it, and because fluctuations of many
times the RMS occur around the sky.  Although a given ATLAS exposure
averages 30 such square degrees, these fluctuations tend to be
spatially correlated and we could potentially incur photometry errors of
as much as 3 times the RMS of the reference catalog.
While it is straightforward to alter APASS DR9 or SkyMapper on a
square degree by square degree basis to agree with Pan-STARRS, this
does not help us for the southern sky.

\subsection{Combining detections}

Initial attempts to assemble the Pathfinder and APASS detections into
a consistent set of photometry over the sky were only successful at the 
10~millimag level --- not good enough for our purposes.  The
reference catalog, photo-flats, and dithers did a good job of making
photometry consistent ($\sim10$ millimag) within a given exposure,
and averaging many detections could then in principle bring the photometry to the
accuracy we need.

However, determination of zeropoints for the exposures to make a
combination whose systematic error was below 10~millimag proved to be
difficult.  We used a combination of three methods to determine zeropoints for exposures:
1) set the zeropoint directly by comparison
with external authority, 2) set the zeropoint by regression against
airmass on photometric nights, and 3) set the zeropoint by
intercomparison of overlapping exposures 
\citep[the ubercal approach;][]{Padmanabhan08,Schlafly12}. 

Although Pan-STARRS is an excellent external authority and is the basis
for the photo-flats,
it does not exist south of Dec $-30$\deg. Gaia is 
available all-sky, but provides only very broad bands and it is a priori 
not clear how well they constrain the narrower Pan-STARRS passbands.
It turned out to be challenging to determine zeropoints from airmass
regression for two reasons.
We were simply reluctant to severely restrict the list of workable nights 
in search of truly photometric nights; we did not feel 
that we would have enough exposures remaining to cover the sky.
The second reason is that zeropoint for an exposure actually depends
on more than airmass.  Since the zeropoint is the conversion between
summed flux and magnitude it depends on the details of photometry
algorithm, not just the aperture and detector QE and gain. In particular, the pixel size of
6.2\arcsec\ (Pathfinder) or 2.6\arcsec\ (APASS) means that 
variations in star density cause very substantial
shifts in the zeropoint, and PSF size and shape are
additional factors.  The zeropoint for images derived by
comparison with Pan-STARRS (0.26\arcsec\ pixels) during a night
depends markedly on galactic latitude and photometry
algorithm and parameters as well as the expected dependency on airmass.

We wrote an elaborate program to simultaneously solve for 
as many as $10^5$ zeropoints based on inter-exposure
differences (the third method) in addition to absolute zeropoints provided by an external
authority (first method) and by airmass regression (second method).
This procedure worked very well.  With external zeropoints weighted
lightly the resulting error in star magnitudes (judged by comparison
with Pan-STARRS in the belt with $-30\deg<\hbox{Dec}<+20\deg$) are
very smooth but have large scale variations as might be expected from
the differential constraint.  When the zeropoints from airmass
regression and external authority were heavily weighted the small scale 
variation became much
rougher on the scale of an observation footprint in the south where we
did not have external zeropoints, but the RMS improved
slightly.

The RMS values of the re-flattened APASS images were
17--12~millimag when we let the inter-exposure differences dominate the
fit weights (labeled diff in the second column of Table 1), and
13--11~millimag when zeropoints from airmass regression dominated
(labeled secz in Table~\ref{tab:zp}).  Optimizing the weights between
differences and external zeropoints made things slightly better still,
but we could not find a combination that made the RMS values
substantially better than 10~millimag.  Since this is the RMS between
square degree tiles it represents an unacceptable systematic error.

\subsection{The Gaia+2MASS+PS1 (GMP) subset and analysis}
\label{sec:GMP}

The release of Gaia DR2 created a new opportunity to set zeropoints.
The Gaia bandpasses are very broad and very different from $griz$.
$G$ is approximately 410--840~nm, $G_{BP}$ is approximately
335--660~nm, and $G_{RP}$ is approximately 640--900~nm, but $G$ and
$G_{RP}$ have extensive red tails, and $G_{BP}$ a substantial dip between
365--400~nm.\footnote{\tt https://www.cosmos.esa.int/web/gaia/dr2}
However, comparison between Gaia magnitudes synthesized from Pan-STARRS
$griz$ revealed little spatial variation across the sky.  It therefore
appeared to be possible to pick a subset of Gaia stars that
match with the 2MASS survey and have low reddening, and use the Gaia
and 2MASS magnitudes to make accurate estimates of $griz$ magnitudes
over the entire sky.  Since our
goal is to set the overall zeropoint for exposures with photometry
for many stars and good internal photometric consistency, a small
subset of the stars can serve to set the zeropoint for all stars in
that exposure.

Experiments revealed that the most fruitful comparison was $griz$ with
Gaia magnitudes in the $(G_{BP}-G_{RP})$, $(J-H)$ plane. The stellar locus
has a tight regression for unreddened, non-red stars, and the
reddening vector is distinctly non-parallel to the stellar locus so
$(J-H)$ can serve to de-redden as well as distinguish giant from dwarf
stars.  Comparing the Gaia and 2MASS data with the Pan-STARRS DR1
catalog revealed an appropriate subset of stars
and regressions for Pan-STARRS $griz$ from the Gaia and 2MASS magnitudes (abbreviated GMP).

The stars that are chosen for the GMP subset have
\begin{itemize}
  \item{} neither Gaia {\tt VARIABLE} nor {\tt DUPLICATE} flag set,
  \item{} uncertainties in $G_{BP}$ and $G_{RP}$ less than 0.03 mag,
  \item{} uncertainties in $J$ and $H$ less than 0.05 mag,
  \item{} $0.7 \le (G_{BP}-G_{RP}) \le 1.4$, \qquad $0.2 \le (J-H)) \le 0.5$, \qquad $AG<1$, and
  \item{} $(G-G_{RP}) - (G-R)_{SSL} < 0.05$,
\end{itemize}
where
\begin{equation}
 (G-R)_{SSL} = +0.0175 + 0.642\,(G_{BP}-G_{RP}) - 0.0784\,(G_{BP}-G_{RP})^2 + 0.002\,(G_{BP}-G_{RP})^3
\label{eq:gaiaslr}
\end{equation}
is the Gaia stellar locus.
Evidently the color cuts are removing very blue and very red stars,
regardless of whether the color is intrinsic or caused by reddening.  We
are also removing stars with $AG\ge1$ that Gaia believes are extremely
reddened
($AG$ is the Gaia extinction estimate in the Gaia broadband $G$ bandpass), 
and the last cut is removing stars that deviate from the
stellar locus in Gaia colors.

We found that the residuals of Pan-STARRS magnitudes with respect to a
regression based on the GMP subset of Gaia and 2MASS magnitudes still had
rather large errors ($\sim$0.1~mag) near the galactic plane, 
but these correlated well with star density.
These deviations are
minimal below a threshold star density, but then increase roughly 
proportional to the square root of star density (inverse of mean
star separation) in more crowded regions.
We use the count of Gaia stars with $15<G<16$ in each angular square
degree ($N_{15-16}$) 
to create a star proximity variable we term
$ng=(N_{15-16})^{1/2}$.  In $i$-band, for example,
the threshold for disagreement is $ng\sim50$ (2500 stars per square degree), and
the disagreement rises to $\sim$100~millimag at $ng\sim100$.

We do not understand the origin of these
deviations.  They do not depend directly on galactic latitude
(extremely obscured areas right on the plane have small $ng$ and small
deviation).  They do not appear to depend on star brightness, they are
still present when 2MASS is removed from the Gaia-Pan-STARRS
comparison, and they do not appear to correlate well with reddening, neither
Gaia estimates nor the total column reddening from \cite{SFD} (SFD).  Star
density correlates with extinction and giant/dwarf ratio, of course,
but neither of these variables correlates as well as star density.
Our best guess is that Gaia incurs
some sort of photometry offset as it scans areas with very high
star densities that creates a small error.

Our predictions for Pan-STARRS $griz$ from Gaia and 2MASS are for the GMP subset are:
\begin{align}\label{eq:gmp}
 (g_{P1} - G_{BP}) &= -0.0321 -0.0520\,X +0.1314\,X^2 +0.0084\,Y +0.00189\,n +0.01326\,n^2 -0.00243\,n^3 \\
 (r_{P1} - G)      &= +0.1940 -0.4013\,X +0.2022\,X^2 -0.0974\,Y +0.02234\,n +0.00352\,n^2 -0.00124\,n^3 \\
 (i_{P1} - G_{RP}) &= +0.3564 -0.0268\,X +0.0429\,X^2 -0.0537\,Y -0.01324\,n +0.03023\,n^2 -0.00484\,n^3 \\
 (z_{P1} - G_{RP}) &= +0.4706 -0.1733\,X -0.0362\,X^2 +0.0714\,Y -0.02992\,n +0.03099\,n^2 -0.00509\,n^3.
\end{align}
where $X=(G_{BP}-G_{RP})$ is the Gaia color, $Y=(J-H)$ is the 2MASS
color, and $n=ng/50=(N_{15-16})^{1/2}/50$ is the scaled, square root star density.

The advantage of using GMP to set zeropoints is its uniformity over
the entire sky.  The cost of such an approach is that errors in the
Gaia and 2MASS source catalogs are impressed on all the results, and
the risk is that systematic changes in reddening, metallicity,
dwarf/giant population ratio, and crowding will also create systematic
biases.  Our judgement is that the Gaia systematic errors are less than 0.01~mag
(although we are not certain about crowded regions), and that the
coefficients multiplying $(J-H)$ are small enough to ensure that the
2MASS systematic errors will also not contribute at the 0.01~mag level
either.  We examine the final results for systematic errors below, and
find they are small.  Using GMP to set the photometric basis for our
reference catalog appears to be the best we can do for photometry
south of $-30$\deg\ Dec, at least until the release of Gaia DR3
which will presumably synthesize $griz$ from the $BP/RP$
spectra.

We created a GMP catalog of stars for the entire sky using the
regressions of Equations~\ref{eq:gmp}-5, and used them to set zeropoints for
all our APASS re-reductions and Pathfinder observations.
Since GMP is a regression that produces Pan-STARRS bandpass magnitudes
from Gaia and 2MASS, we must first convert the Pan-STARRS $g_{P1}$,
$r_{P1}$, $i_{P1}$, and $z_{P1}$ to each individual catalog's bandpass
before comparison.  To our knowledge, only Pan-STARRS has published
bandpasses measured in-situ, so we did not trust any
estimates of color terms derived from integration of SEDs
against bandpass estimates, but worked entirely empirically.
Regressions for catalog magnitude as a function of
Pan-STARRS magnitude and vice versa over the usual Dec band are given in
Table~\ref{tab:cterm}.

The relations in Table~\ref{tab:cterm} are {\it not} inverses.  The calculation of catalog
magnitude as a function of Pan-STARRS was derived from the GMP subset
of stars in order to set zeropoints (the same relation is applicable
to both the APASS re-reduction as well as APASS DR9.)  The relations
that predict Pan-STARRS magnitude as a function of catalog magnitude
conversely are calculated from {\it all} stars, regardless of color or
reddening.  In order to derive these, we matched all stars from each
catalog to Pan-STARRS, sorted the independent variable $(g-r)$ or
$(r-i)$ from the catalog into bins of size 0.1 mag, and then took the
median of all magnitude differences in that bin.  We fitted a line to
a selected range in abscissa and verified that it matched trends
visible in these medians as well as the cloud of all points.  The
results for the APASS re-reduction and APASS DR9 are quite independent
of each other because the APASS re-reduction includes much fainter and
redder stars than APASS DR9.  As with all simple, empirical, linear
color term transformations, these do not pretend to high accuracy for
stars with extremely red or blue SEDs.  Therefore individual \refcat\ magnitudes
for very red or blue stars without direct Pan-STARRS photometry (south of 
Dec $-30$\deg\ or brighter than $m\sim14$) should be treated with care.

\begin{table}[htp]
\caption{Color terms}
\begin{center}
\begin{tabular}{ccrrr}
\hline
   $y$          & $x$            &  $C_0$   & $C_1$    &  RMS  \\
\hline                    
$(g_{AP}-g_{P1})$&$(g_{P1}-r_{P1})$& $ 0.023$ & $ 0.054$ & 0.032 \\
$(r_{AP}-r_{P1})$&$(g_{P1}-r_{P1})$& $-0.058$ & $ 0.023$ & 0.039 \\
$(i_{AP}-i_{P1})$&$(g_{P1}-r_{P1})$& $ 0.003$ & $ 0.057$ & 0.050 \\
$(g_{PT}-g_{P1})$&$(g_{P1}-r_{P1})$& $-0.007$ & $ 0.045$ & 0.043 \\ 
$(r_{PT}-r_{P1})$&$(g_{P1}-r_{P1})$& $ 0.018$ & $-0.034$ & 0.036 \\
$(i_{PT}-i_{P1})$&$(g_{P1}-r_{P1})$& $ 0.020$ & $-0.029$ & 0.042 \\
$(g_{SM}-g_{P1})$&$(g_{P1}-r_{P1})$& $ 0.010$ & $-0.228$ & 0.032 \\ 
$(r_{SM}-r_{P1})$&$(g_{P1}-r_{P1})$& $ 0.004$ & $ 0.039$ & 0.016 \\
$(i_{SM}-i_{P1})$&$(r_{P1}-i_{P1})$& $ 0.008$ & $-0.110$ & 0.022 \\
$(z_{SM}-z_{P1})$&$(r_{P1}-i_{P1})$& $-0.004$ & $-0.097$ & 0.020 \\
\hline                    
$(g_{P1}-g_{AP})$&$(g_{AP}-r_{AP})$& $-0.009$ & $-0.061$ & 0.026 \\
$(r_{P1}-r_{AP})$&$(g_{AP}-r_{AP})$& $ 0.065$ & $-0.026$ & 0.027 \\
$(i_{P1}-i_{AP})$&$(r_{AP}-i_{AP})$& $-0.015$ & $-0.068$ & 0.045 \\
$(g_{P1}-g_{A9})$&$(g_{A9}-r_{A9})$& $-0.008$ & $-0.099$ & 0.021 \\
$(r_{P1}-r_{A9})$&$(g_{A9}-r_{A9})$& $-0.050$ & $ 0.040$ & 0.026 \\
$(i_{P1}-i_{A9})$&$(r_{A9}-i_{A9})$& $-0.015$ & $ 0.042$ & 0.046 \\
$(g_{P1}-g_{PT})$&$(g_{PT}-r_{PT})$& $ 0.012$ & $-0.048$ & 0.018 \\
$(r_{P1}-r_{PT})$&$(g_{PT}-r_{PT})$& $-0.017$ & $ 0.035$ & 0.019 \\
$(i_{P1}-i_{PT})$&$(r_{PT}-i_{PT})$& $-0.011$ & $ 0.053$ & 0.025 \\
$(g_{P1}-g_{SM})$&$(g_{SM}-r_{SM})$& $ 0.004$ & $ 0.272$ & 0.029 \\
$(r_{P1}-r_{SM})$&$(g_{SM}-r_{SM})$& $-0.016$ & $-0.035$ & 0.021 \\
$(i_{P1}-i_{SM})$&$(r_{SM}-i_{SM})$& $-0.011$ & $ 0.100$ & 0.016 \\
$(z_{P1}-z_{SM})$&$(r_{SM}-i_{SM})$& $ 0.009$ & $ 0.082$ & 0.020 \\
\hline
\end{tabular}
\end{center}
\tablecomments{Polynomial coefficients $y=C_0+C_1\,x$ are listed for
  conversions between catalog and Pan-STARRS magnitudes. The subscripts are as follows : 
  Pan-STARRS ($P1$), ATLAS Pathfinder ($PT$), Skymapper ($SM$), reprocessed APASS ($AP$), and APASS DR9 ($A9$) .}
\label{tab:cterm}
\end{table}%

The APASS re-reduction and Pathfinder detections were then grouped by
matching against the list of all Gaia stars and weighted median
magnitudes were calculated for each star.  (The ``weighted median''
occurs at the 50\% quantile of the cumulative inverse variance weights
for all the detection magnitudes.)  An uncertainty was estimated from
the weighted quartiles and the number of contributing observations.
When this uncertainty is significantly discrepant from the individual
uncertainties we flag the star as a potential variable.

There are real stars in the APASS and Pathfinder observations for which
the closest Gaia star is more than 10.8\arcsec\ distant.
For these we assembled a median magnitude by
grouping among themselves, and we include them in the \refcat\ output.
We endeavored to avoid galaxies in the {\tt DoPhot} output for
APASS and Pathfinder, but these images do not have a lot of spatial
resolution so a few bright galaxies have crept into \refcat\ (about 0.02\% of the total).
These non-Gaia objects can be distinguished because the Gaia magnitude
and uncertainty are zero.  Appendix~\ref{app:details} provides details.

The photometry of the final APASS and Pathfinder objects are compared against
Pan-STARRS in Table~\ref{tab:zp} with GMP listed as the
source of zeropoint.  The scatter for APASS and Pathfinder are now
only a few millimags.

We also used the GMP set of stars to correct the photometry of APASS
DR9.  Because the reduction, photometric analysis,
and detection combination used by the APASS project for DR9 is
completely different from our re-reduction, there is merit in treating
it as an independent set of photometry.  We
cannot re-flatten the APASS DR9 observations and recombine them, but we can
correct on a square degree basis using the GMP subset, and as noted
above much of the variation of APASS DR9 with respect to Pan-STARRS
occurs on scales of many degrees, so a square degree correction should
not create serious discontinuities.

Similarly we created a new version of SkyMapper DR1.1 where the
magnitudes for $griz$ in each square degree are shifted to agree
with the GMP estimates.  SkyMapper DR1.1 flags
some stars as having uncertain photometry because of crowding. 
For these we added 0.2 magnitude in quadrature to
the uncertainties.  This 0.2 mag is really an operational choice rather
than an estimate of true uncertainty.  The effect causes their contribution
to the final magnitude to be minimal if another catalog has data, and
serves to flag uncertain photometry when only SkyMapper contributes.

For both APASS DR9 and SkyMapper the magnitudes in each filter and
each square degree are shifted by a constant value, so relative photometry
between stars of different brightness in a given square degree 
is not changed.  The regressions to
convert these revised APASS DR9 and SkyMapper magnitude to the
Pan-STARRS bandpasses are given in Table~\ref{tab:cterm}.
 
Lastly we list in Table~\ref{tab:gcterm} a crude set of regressions to
get Pan-STARRS magnitude estimates from Gaia magnitudes alone in order to
provide some sort of magnitude estimate when no other catalog is available.  
(These are given large systematic error which de-weights their contribution 
if some other catalog provides real $griz$ photometry.)
There is a large
scatter, of course, and these are not available for stars too faint to
have $G_{BP}$ and $G_{RP}$.  There is a prominent set of red stars
(M giants) that diverge significantly from this relationship, but for
magnitudes fainter than $m>14$ the M dwarfs will dominate.  
\begin{table}[htp]
\caption{Gaia-only color terms}
\begin{center}
\begin{tabular}{crrl}
\hline
   $y$           & $C_0$ & $C_1$ & Domain \\
\hline                    
$(g_{P1}-G_{RP})$ & $-0.19$ & $ 1.25$ & $<1.5$ \\
$(g_{P1}-G_{RP})$ & $-0.50$ & $ 1.45$ & $\ge1.5$ \\
$(r_{P1}-G_{RP})$ & $ 0.08$ & $ 0.48$ & $<1.5$ \\
$(r_{P1}-G_{RP})$ & $-0.23$ & $ 0.69$ & $\ge1.5$ \\
$(i_{P1}-G_{RP})$ & $ 0.34$ & $ 0.03$ & $<1.5$ \\
$(i_{P1}-G_{RP})$ & $ 0.17$ & $ 0.14$ & $\ge1.5$ \\
$(z_{P1}-G_{RP})$ & $ 0.49$ & $-0.21$ & $<1.5$ \\
$(z_{P1}-G_{RP})$ & $ 0.45$ & $-0.18$ & $\ge1.5$ \\
\hline
\end{tabular}
\end{center}
\tablecomments{Polynomial coefficients $y=C_0+C_1\,(G_{BP}-G_{RP})$
  to convert Gaia magnitudes to Pan-STARRS magnitudes.  The broken
  linear relations change coefficients according to the value of 
  $(G_{BP}-G_{RP})$ and the ``Domain'' column.}
\label{tab:gcterm}
\end{table}%

\section{Catalog combination}

The pieces of our reference catalog now need to be combined.  For the
master list of stars we collect all Gaia stars for which at least one
of $G$, $G_{BP}$, or $G_{RP}$ is brighter than $m\le19$.  To this list
we add objects from any of the other catalogs which seem
to be bona fide stars and which are farther than 0.001\deg\ (3.6\arcsec)
from the nearest Gaia star (0.003\deg\ for Pathfinder and APASS).  
By including stars from Tycho-2 and the BSC we think \refcat\ contains
all stars brighter than $m\sim10$ with no omissions, although the usual
caveats about unresolved binary stars and faint stars near bright stars 
apply.  As noted above, we do not
expect stars in \refcat\ with $m<10$ to have particularly accurate
photometry although the coordinates should be quite good.  These non-Gaia
stars are a very small fraction ($\sim0.1$\%) of \refcat.
Details are provided in Appendix~\ref{app:details}.

Matching between catalogs is done with a tolerance of 0.0002\deg\ (0.72\arcsec).
Pathfinder and APASS use Gaia DR1 coordinates (epoch 2015.5), and Tycho and BSC
are shifted to epoch 2015.5.  Pan-STARRS DR1 is epoch $\sim$2013, but we did not
shift Gaia coordinates to that (imprecise) epoch so a handful of high proper
motion stars or stars with a very large parallax may have failed to match
correctly.  SkyMapper coordinates derive from UCAC4 (epoch 2000) so match failures
will occur for stars with proper motion larger than $\sim$50~milliarcsec/yr,
and similarly for APASS DR9.

For this master list we have Gaia data for most stars, 2MASS data for
many stars, and $griz$ photometry for all stars from a variety of
sources, including Gaia, GMP, Pan-STARRS, SkyMapper, Pathfinder, APASS, APASS DR9,
and Tycho-2/BSC, all put on the Pan-STARRS bandpasses
using the relations above.  We soften each star's formal error of each magnitude
by augmenting it in quadrature with a value that reflects what we consider each
catalog's systematic reliability to be: 0.1 for Gaia DR2 only (Table~\ref{tab:gcterm}), 0.015 using
the GMP regression but increasing when star
colors leave the restricted GMP domain, 0.01 for Pan-STARRS but
increasing by 1 mag/mag for $g<14.0$, $r<13.7$, $i<13.5$, and $z<13.2$
and 0.1 mag/deg south of Dec $-30$\deg, 0.02 for SkyMapper, Pathfinder, 
and APASS, 0.05 mag for APASS DR9, and 0.1 magnitude for Tycho-2/BSC.
The estimation of $g_{P1}$, $r_{P1}$, $i_{P1}$, and $z_{P1}$ for each
star is a straightforward process that converts each catalog's data to
Pan-STARRS bandpasses according to Tables \ref{tab:cterm} and \ref{tab:gcterm},
computes the weighted average for each star from
all contributing catalogs, tests whether any contributor is worse than
3$\sigma$ and if so increases the error on the worst offender by
twice its $\sigma$ deviation, and iterating until no contributor
deviates by more than 3$\sigma$.  A bitmap is kept of all contributing
catalogs to each final magnitude as well as a count of the number of times a
contributor had its error increased.

In order to make \refcat\ more useful for observations that do not
enjoy the spatial resolution of Pan-STARRS or Gaia, for each star in
the catalog we calculate the distance to stars in the full Gaia DR2
that are bright enough to interfere with its photometry.  We assemble the
cumulative flux of neighboring stars in Gaia $G$ as a function of distance from each star
and note the radius where this flux exceeds 0.1, 1.0, and 10 times
the central star's $G$ brightness, out to a maximum distance of
36\arcsec.

We also provide the total column extinction values at the location of
each star from \cite{SFD}.  Although many stars will lie in front of some
or all of the dust, this may prove to be helpful for some applications.

More details about the precise content of the catalog may be found in
Appendix~\ref{app:details}.

\section{Catalog properties}

\subsection{Star counts}

\begin{figure}[htbp]
\begin{center}
\centerline{\includegraphics[width=6.5in]{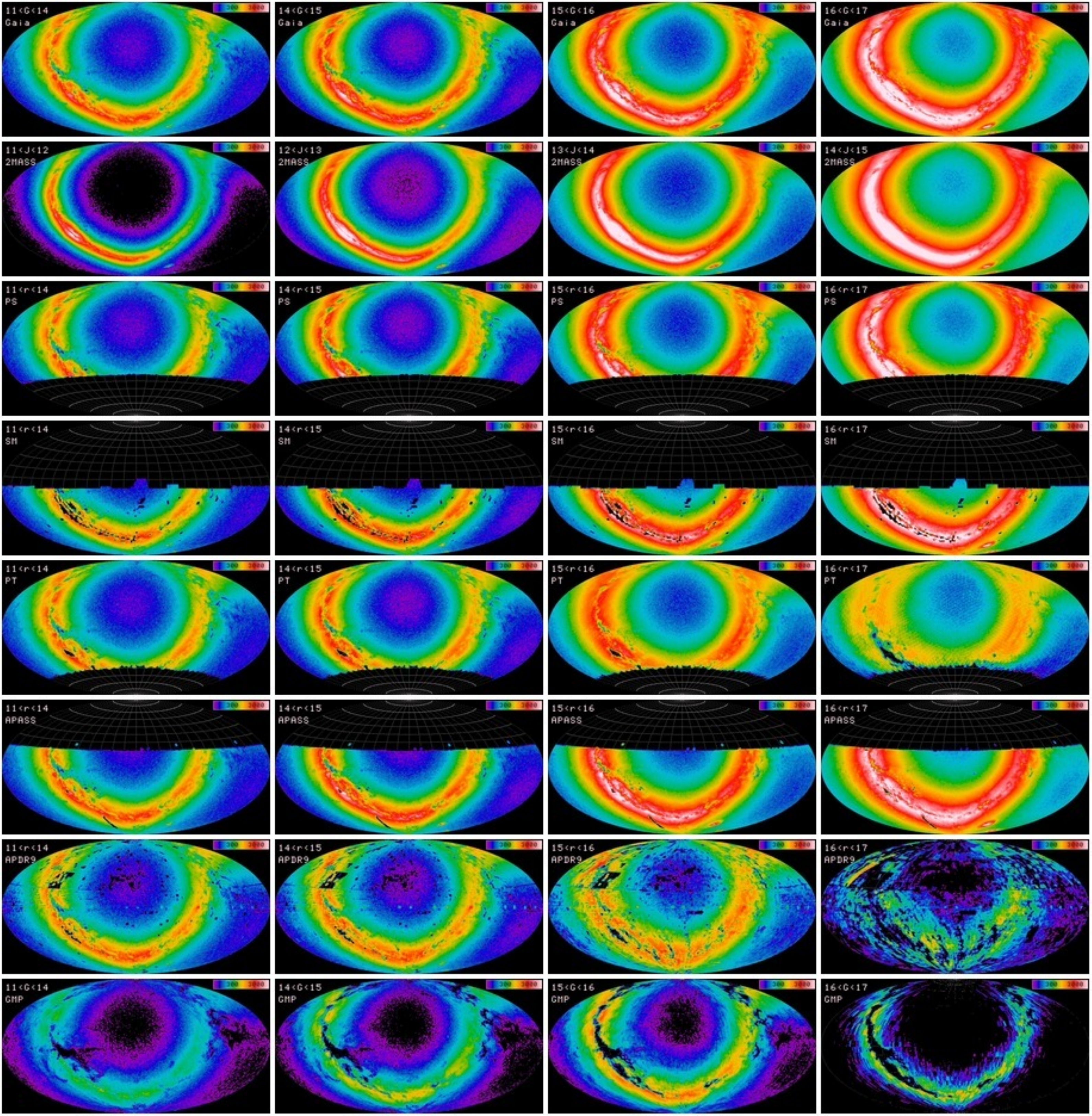}}
\caption{The coverage of the sky by the various contributing surveys
  is shown with color coding the log number of stars per square
  degree.  From top to bottom are Gaia, 2MASS, Pan-STARRS, 
  SkyMapper, Pathfinder, APASS, APASS DR9, and GMP, and from left to right is
  $11<r<14$, $14<r<15$, $15<r<16$, $16<r<17$, except $G$ is used for Gaia
  and the ranges apply to $J+2$ for 2MASS.
  The color scale is logarithmic from 50 to 10000,
  magenta is 70, green is 300, yellow is 700, red is 3000 stars per
  square degree.  The \refcat\ counts are indistinguishable from
  the Gaia counts in the top panel.}
  \label{fig:counts}
\end{center}
\end{figure}

Figure~\ref{fig:counts} shows the star counts in all the contributing
surveys as a function of star brightness.  The assembled \refcat\ looks
no different than the Gaia counts because it includes all Gaia stars
as well as a tiny fraction more.  Many features are
apparent, such as the
progressive failure of Pathfinder to be able to discern crowded stars
fainter than $r>15$, the dramatically lower density of stars in GMP
resulting from the severe selection filters applied, 
the tiny ``crack in the sky'' in the APASS counts, and of course the
markedly different appearance of the galactic plane in the 2MASS $J$-band.

Note that the APASS DR9 counts also show the APASS DR8 stars in the
northern hemisphere, although these were not used in the
production of \refcat.  Pan-STARRS loses stars in $r$-band
relative to the redder Gaia $G$-band due to extinction, but the
correspondence between the stars lists is nearly perfect at these
bright magnitudes. The APASS counts are also very similar to Gaia
for $r<16$, although the completeness for $16<g<17$ begins to fall
off.

\subsection{Catalog comparisons with \refcat}

\begin{figure}[htbp]
\begin{center}
\centerline{\includegraphics[width=7.0in]{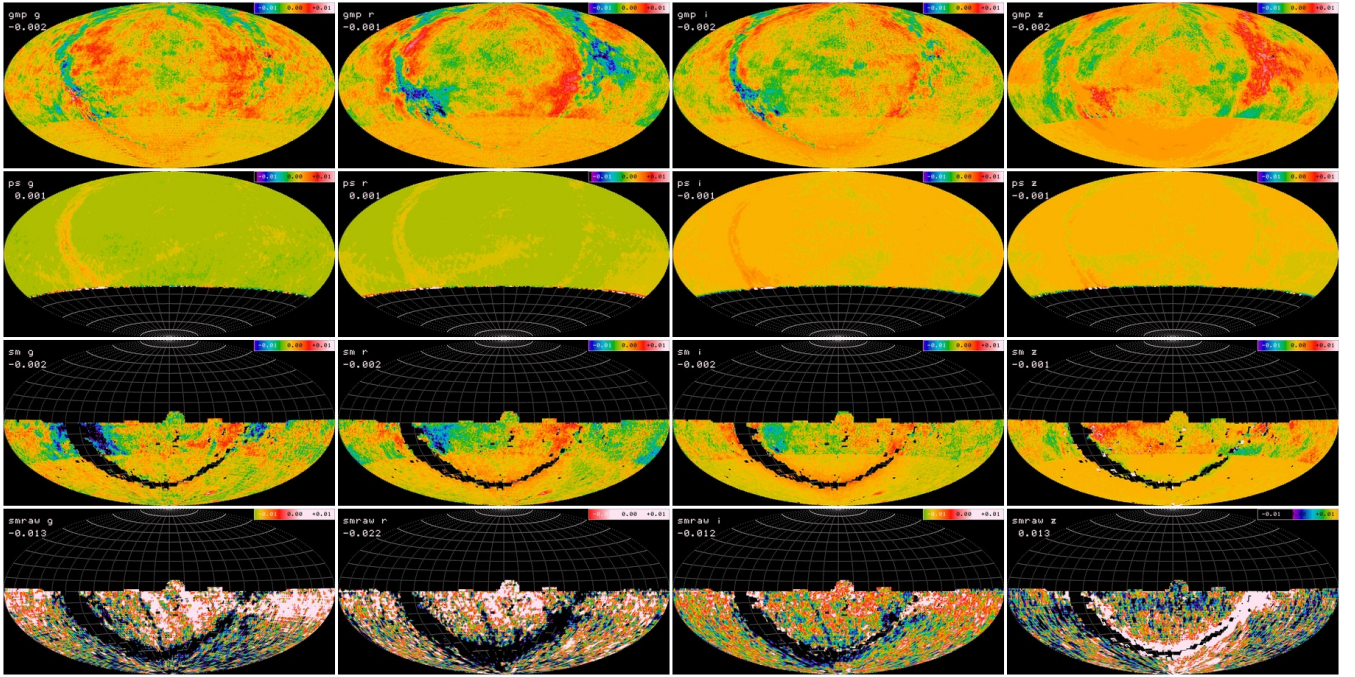}}
\caption{The median difference in $griz$ between star magnitudes in the GMP,
  Pan-STARRS, SkyMapper DR1.1, and uncorrected SkyMapper DR1.1
  catalogs relative to \refcat\ are coded by color,
  evaluated on each square degree in the sky.  The color indicates
  the disagreement between catalog and \refcat: black and white when
  outside the range $\pm15$~millimag and green to red for $\pm5$~millimag.
  A mean offset indicated in the upper left is also subtracted from
  the catalog before comparison with \refcat.  Left to right
  are $g$, $r$, $i$, and $z$ magnitudes, modified from catalog to
  Pan-STARRS according to the color terms of Table~\ref{tab:cterm}
  before subtraction.}
  \label{fig:psmg}
\end{center}
\end{figure}

\begin{figure}[htbp]
\begin{center}
\centerline{\includegraphics[width=5.25in]{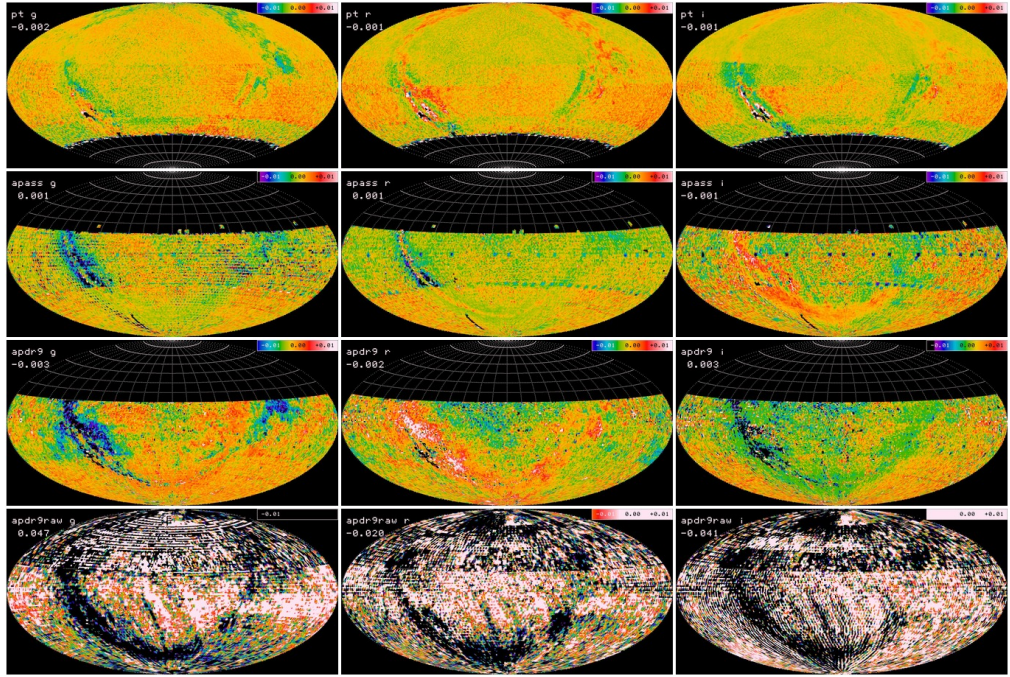}}
\caption{The median difference in $gri$ between star magnitudes in the
  Pathfinder, APASS, APASS DR9, and uncorrected APASS DR9 catalogs
  relative to \refcat\ are coded by color,
  evaluated on each square degree in the sky.  The color indicates
  the disagreement between catalog and \refcat: black and white when
  outside the range $\pm15$~millimag and green to red for $\pm5$~millimag.
  A mean offset indicated in the upper left is also subtracted from
  the catalog before comparison with \refcat.  Left to right
  are $g$, $r$, and $i$ magnitudes, modified from catalog to
  Pan-STARRS according to the color terms of Table~\ref{tab:cterm}
  before subtraction.}
  \label{fig:ptapass}
\end{center}
\end{figure}

Figures~\ref{fig:psmg} and \ref{fig:ptapass} show the difference
between the star magnitudes in each of eight catalogs relative to the
final \refcat.  These are computed as the median difference between
catalog's star magnitudes (converted to Pan-STARRS bandpasses using the
color terms of Table~\ref{tab:cterm}) and \refcat, evaluated on
each square degree of the sky.  Pathfinder and APASS have $g$, $r$,
and $i$ comparisons, and GMP, Pan-STARRS, and SkyMapper also have $z$
comparisons.  The color palette stretches between $-0.015$ and $+0.015$
magnitudes.

Close inspection of the GMP--\refcat\ comparison does not show any obvious Gaia scan
pattern (although it is obvious if the {\tt DUPLICATE} stars are not eliminated), but
it is possible to discern tiny errors
in Pan-STARRS: the 3\deg\ honeycomb pattern and 10\deg\
square chunk pattern can be seen at the 5~millimag level.  Because of the
heavy weighting given to Pan-STARRS these artifacts are not seen in
the Pan-STARRS--\refcat\ comparison, indicating these features have been
carried into \refcat.

The third row of Figure~\ref{fig:psmg}
shows the DR1.1 SkyMapper magnitudes, spatially corrected on a degree scale
to GMP.  The roughness
in the $-30<\hbox{Dec}<+0$ overlap region with Pan-STARRS is relatively
large because its zeropoint is corrected by GMP, not Pan-STARRS. 
Some minor offsets from crowding are visible in the Large Magellanic Cloud.
The fourth row shows the 
uncorrected SkyMapper DR1.1 comparison as a reference
(but we do not use uncorrected magnitudes). 
The flaws that SkyMapper inherited by using zeropoints from
APASS DR9 photometry are visible in this panel.

Figure~\ref{fig:ptapass} shows the comparison with Pathfinder, APASS
as re-reduced for this work, and APASS DR9 published by the APASS
project.  The Pathfinder comparison looks a bit better than GMP where
it overlaps with Pan-STARRS, presumably because the bandpasses are
more closely matched
(compare the top rows of Figures~~\ref{fig:psmg} and \ref{fig:ptapass}). 
The mild discontinuity at Dec
$-30$\deg\ shows the onset of southern systematic error in \refcat.
Pathfinder also reveals a honeycomb pattern on 5\deg\ scales
at the 5~millimag RMS level arising from the Takahashi telescope
vignetting and imperfect flattening.

The APASS comparison reveals the APASS footprint and the
imperfections in the re-flattening we imposed.  The APASS DR9
comparison is based on the magnitudes published by the APASS project
corrected on a square degree basis by GMP, and the comparison looks
somewhat different than the re-reduction of APASS above.  Apart from the
effects from our re-reduction of the images, the APASS DR9 is more than
one magnitude shallower than the APASS re-reduction, so different
stars are being compared.  Some of the degree-by-degree roughness in
these images stems from fluctuations in degrees with few matching
stars.  The fourth row in Figure~\ref{fig:ptapass}  shows the APASS DR9 (and DR8 for Dec north of
$+20$\deg) before correction by GMP.  Not only are there large
differences on degree and much larger scales, the mean offsets listed
in the upper left are significant.

\subsection{HST standards}

The SkyMapper project integrated the Calspec set of HST standard star
SEDs against the Pan-STARRS bandpasses to make
predictions for magnitudes in $g_{P1}$, $r_{P1}$, $i_{P1}$, and
$z_{P1}$-bands.\footnote{\tt
  http://skymapper.anu.edu.au/filter-transformations/}
We do not expect
great accuracy for \refcat\ magnitudes brighter than $m<10$ because the
only contributions are from Tycho-2, BSC, or Gaia,
with rather gross color transformations to $griz$.

\begin{figure}
\begin{center}$
\begin{array}{ccc}
\includegraphics[width=3in]{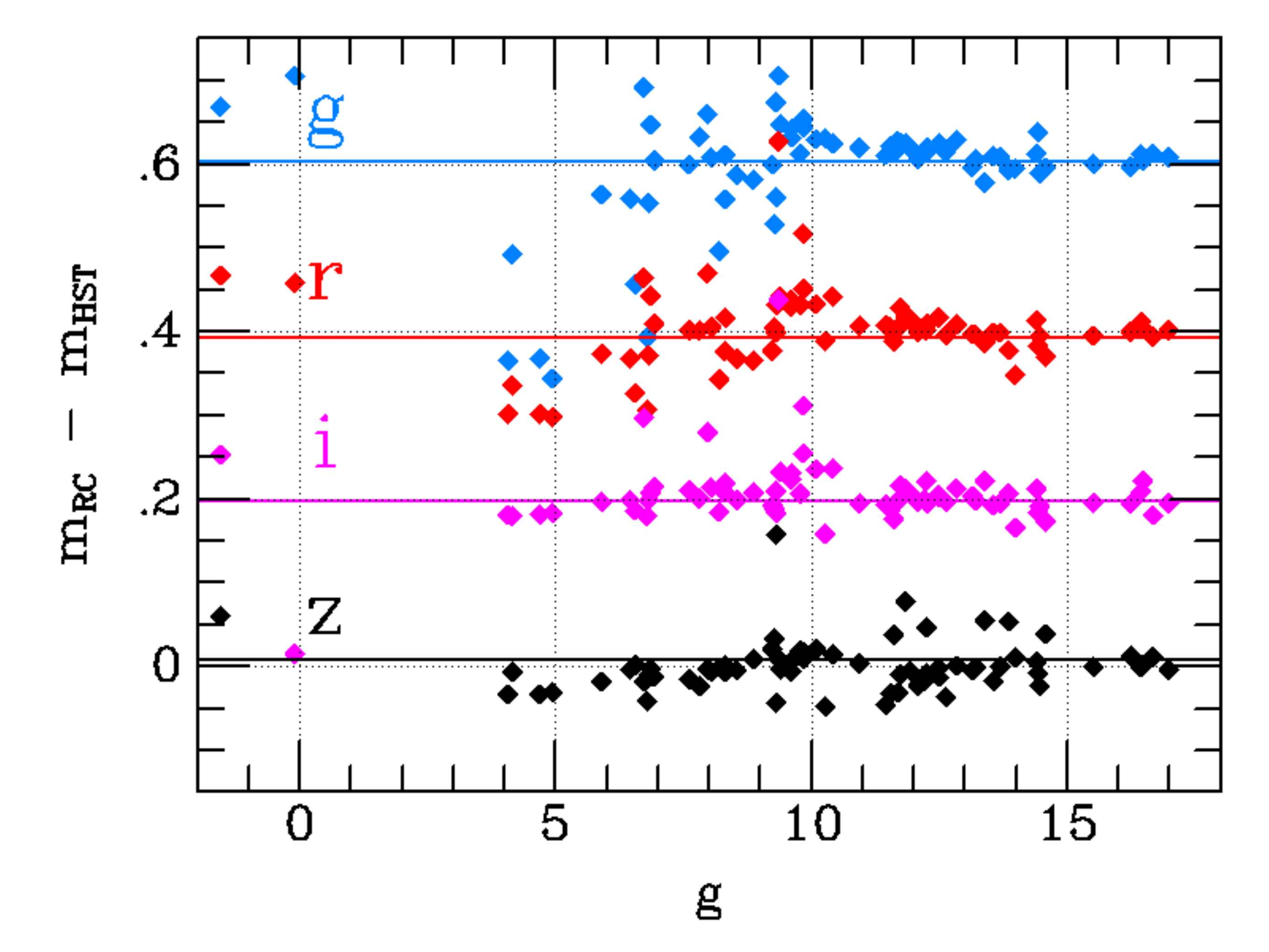}
\qquad
\includegraphics[width=3in]{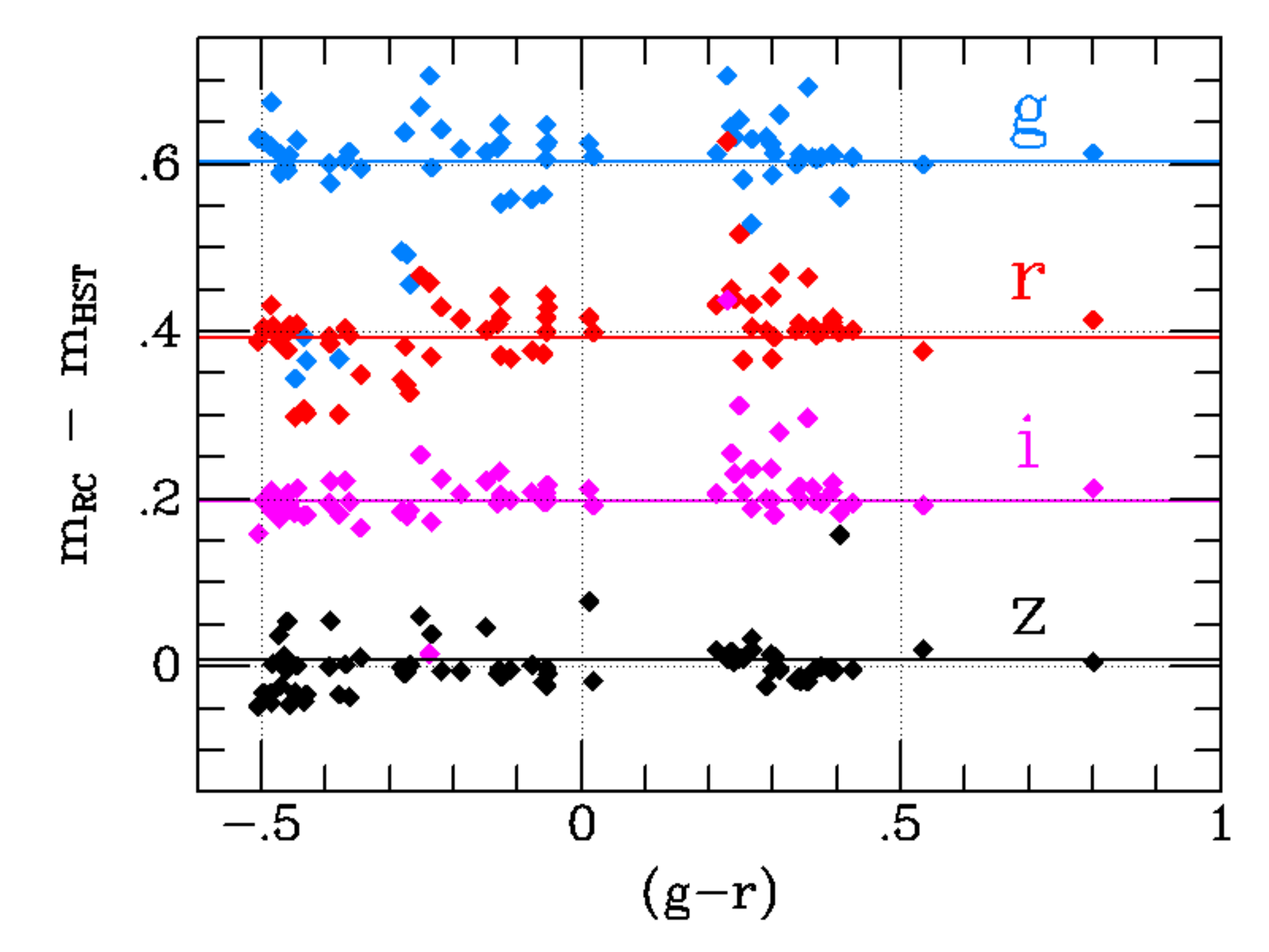}
\end{array}$
\end{center}
\caption{The comparison between \refcat\ $griz$ magnitudes and those
  integrated from 70 HST Calspec SEDs over the Pan-STARRS bandpasses
  are shown, offset by 0.2 magnitudes for clarity.  Left is the
  difference as a function of $g$, right is $(g-r)$.  Three regimes of
  accuracy ($m<10$ from Tycho-2, $10<m<14$ from Pathfinder and APASS, and
  $m>14$ from Pan-STARRS) are evident.}
\label{fig:hst}
\end{figure}

The difference between \refcat\ magnitude and HST Calspec magnitudes
are shown in Figure~\ref{fig:hst} as a function of $g$ magnitude and
$(g-r)$ color.
Restricting to stars fainter than $g>10$ we find the difference
between \refcat\ and 36 HST standards to be
$\Delta g = +0.014\pm0.013$,
$\Delta r = +0.000\pm0.016$,
$\Delta i = -0.001\pm0.017$, and
$\Delta z = +0.000\pm0.025$.
Restricting to stars fainter than $g>13$ the differences with respect
to 17 HST standards is
$\Delta g = +0.003\pm0.013$,
$\Delta r = -0.007\pm0.015$,
$\Delta i = -0.003\pm0.015$, and
$\Delta z = +0.008\pm0.021$.
We do not regard any of these differences to be significant enough to
compel us to adjust the absolute zeropoint that has been set by GMP
and its calibration against Pan-STARRS.  There are 8 HST standards
south of Dec $-30$\deg, but only 1 is fainter than $g>10$.  These
scatter evenly among the northern stars and do not show any offset in
any of these bandpasses, confirming that
\refcat\ maintains all-sky accuracy even where Pan-STARRS
makes no contribution.

\subsection{\refcat\ errors and uncertainties}

The $griz$ photometry in \refcat\ has both statistical and systematic
errors.  The statistical errors listed for each star's magnitude in \refcat\ 
are derived from the errors of extant
catalogs and photometry, augmented by judicious systematic allowances
according to the trustworthiness of zeropoint adjustments we have applied
as well as the degree of color correction from catalog to standard 
Pan-STARRS bandpasses.  The combination process of
photometry from different catalogs appears to obey Gaussian statistics
for most stars, which gives us confidence that the individual star magnitudes
have reasonable error estimates.  The distribution of $\chi^2$/{\tt DOF} and 
the frequency of outlier rejection provided
in the catalog are consistent with a Gaussian distribution with a few
percent tail from variable stars and other outliers.  However \refcat\ remains a heterogeneous
compilation at some level, and examination of magnitude error as a function
of magnitude will reveal disjoint clumps according to the dominant 
contributor during the catalog combination.

It is difficult to be certain about systematic error and
fidelity to absolute photometry.  The comparison with the STScI Calspec
standards affirms the absolute photometry heritage of \refcat\ through
Pan-STARRS.  Comparing Gaia with Pan-STARRS, two completely different
sets of photometry with completely different processes to achieve internal
consistency, is a means to understand the systematic errors that may be
present in each.

The top two panels in Figure~\ref{fig:psmg} 
illustrate the difference between GMP versus \refcat\ and Pan-STARRS versus \refcat.
The difference between Pan-STARRS and \refcat\
in the second panel is virtually non-existent, as expected because Pan-STARRS observations
have lower error than any other contributor to \refcat\ and therefore
strongly dominate the weighted average magnitudes for stars.

However, GMP is also fundamental to \refcat\
because it provides the normalization for the Pathfinder and APASS
exposures that were re-reduced as well as the re-normalization of
APASS DR9 and SkyMapper DR1.1.  Therefore, by construction the difference
between GMP and \refcat\ south of Dec $-30$\deg\ is also very small.  

\begin{figure}[htbp]
\begin{center}$
\begin{array}{ccc}
\includegraphics[width=3in]{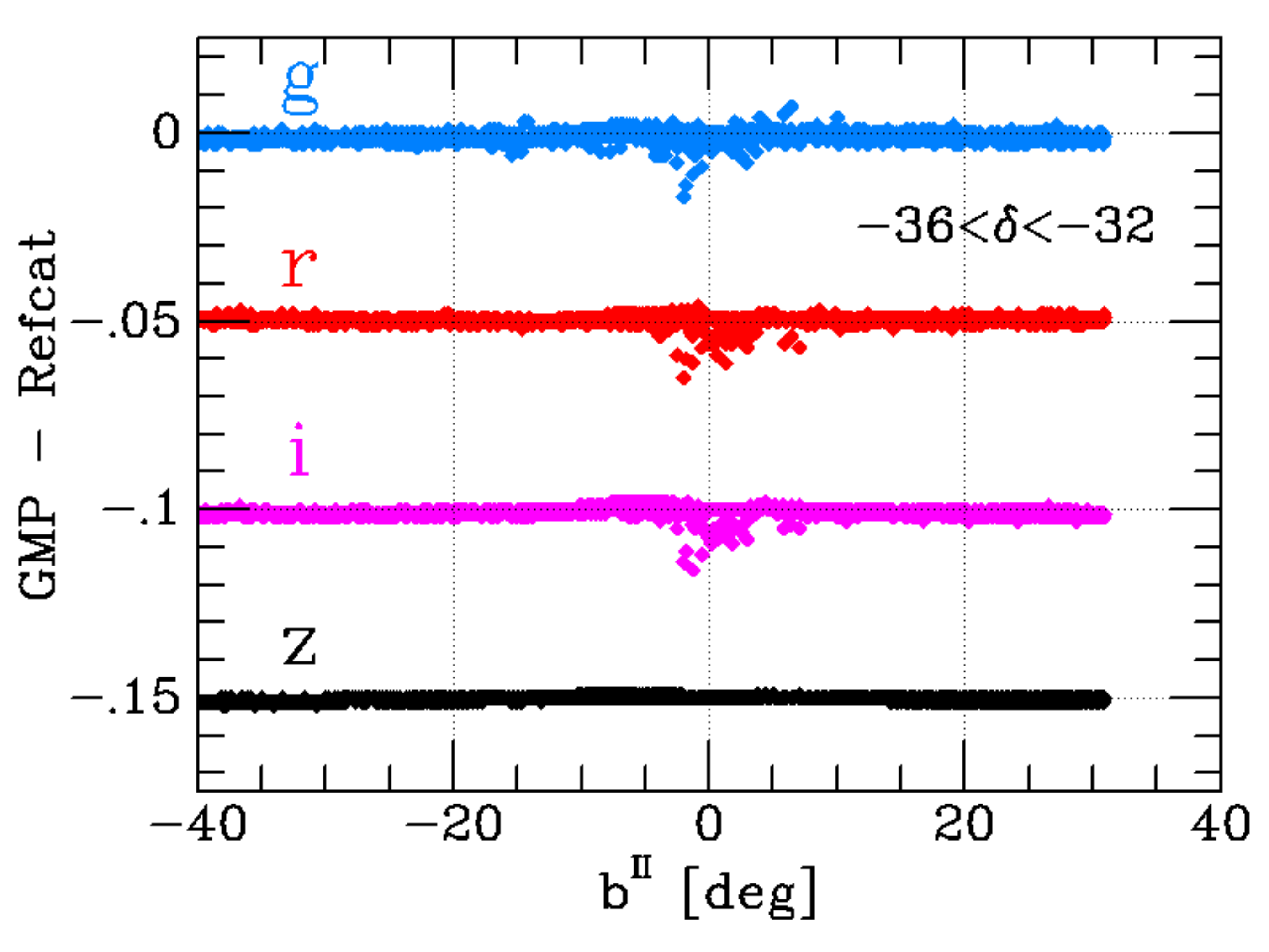}
\qquad
\includegraphics[width=3in]{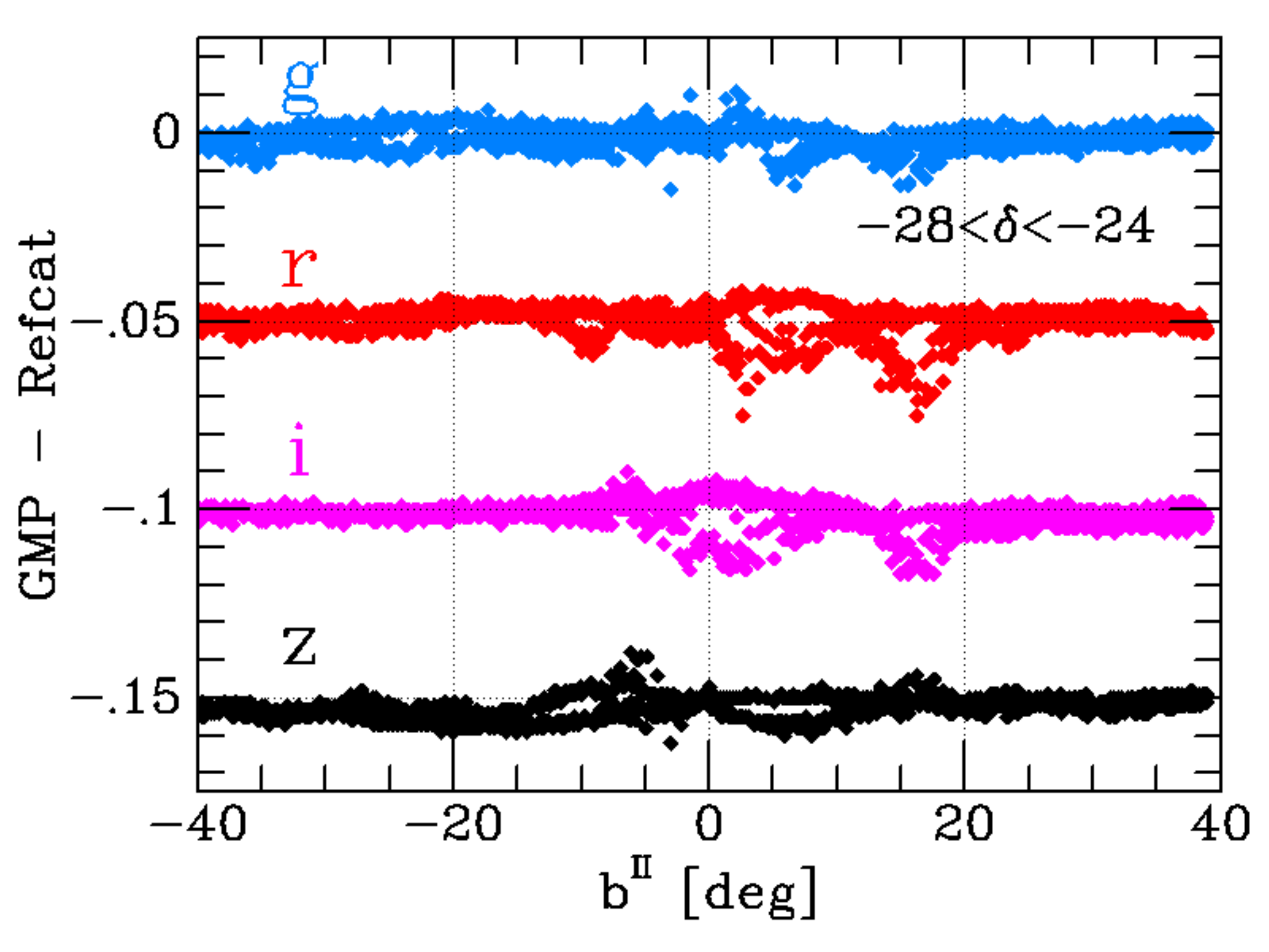}
\end{array}$
\end{center}
\caption{The median differences 
  between $griz$ star magnitudes in the GMP
  catalog relative to \refcat\ evaluated on square degrees (shifted by 0.05 mag for
  clarity) are shown as a function of galactic
  latitude for two bands: $-36<\hbox{Dec}<-32$ (left) and
  $-28<\hbox{Dec}<-24$ (right).  The northern points (right)
  display differences in GMP relative to \refcat\ and by extension
  Pan-STARRS; the southern points presumably have similar errors
  with respect to Pan-STARRS, but their comparison with \refcat\
  is very nearly zero because GMP set the zeropoints of all the
  observations contributing to \refcat\ south of $-30$\deg.}
  \label{fig:gmpref}
\end{figure}

We can therefore get a sense of the systematic inaccuracy of
\refcat\ by looking at the discontinuity in
the GMP--\refcat\ comparison north and south of Dec $-30$\deg.  
Figure~\ref{fig:gmpref} shows the
GMP--\refcat\ comparison evaluated on square degrees 
in a pair of 4\deg\ wide strips 
immediately above and below the discontinuity.

The northern strip shows
marked deviations from zero, exceeding $\pm20$~millimag at 
particular locations although
the RMS is much less than 10~millimag.  $r$ band has the
worst residuals: 
5119 (coordinate) square degrees deviate by more than 5~millimag,
771 square degrees deviate by more than 10~millimag, and
23 square degrees deviate by more than 20~millimag.  $g$ band
never deviates by as much as 20~millimag, $i$ band deviates by
at least 20~millimag on 5 square degrees, and $z$ band on 1
square degree.
Formally the difference
between GMP and Pan-STARRS evaluated on each square degree over the
entire sky north of Dec $-28$\deg\ has an RMS of about 2~millimag but of
course it is the excursions that are of concern.  Because the
Pan-STARRS--\refcat\ comparison has no such deviations we can
conclude that these reflects error in the GMP predictions for
Pan-STARRS magnitudes.
We presume that GMP is continuous and
actually has errors akin to the ones visible in the north strip, and
so \refcat\ has such errors south of Dec $-30$\deg.  As noted above,
the GMP--\refcat\ comparison north of Dec $-30$\deg\ reveals hints
of the Pan-STARRS footprint and chunks, so the northern part of
\refcat\ surely also has systematic errors of at least 2~millimag.

Although the all-sky RMS is about 2~millimag, there are obvious
coherent patches, particularly in $r$ and $z$ near the galactic anticenter
where the GMP--\refcat\ difference is consistently around $\pm5$~millimag.
These patches remain from the density correction terms of Equation~\ref{eq:gmp}
that erased much bigger errors near the galactic plane.  Although we have a suspicion
that this has to do with Gaia--Pan-STARRS filter mismatch and gradients
in stellar population or metallicity, we were unable to
find an all-sky correlate with these features.  It is also conceivable
that the ubercal procedure that sets the internal consistency of the
Pan-STARRS photometry has wandered because of poor constraints
on large scale error.  However, for our purposes this level of error is
not important and we do not have an all-sky method to correct for it.

Therefore our summary assessment of the accuracy of \refcat\ is that
by construction it matches Pan-STARRS north of Dec $-30$\deg\ with
an RMS systematic error of at least 2~millimag, and
south of Dec $-30$\deg\ it has an RMS systematic error of at least
3~millimag, except near the galactic center where systematic
errors can occasionally grow to more than~20 millimag.  However,
note that these excursions are more correlated with stellar density
than galactic latitude; the differences are small right on the
plane because the extinction is so high that the star density is lower.

\section{Summary and Conclusions}

We have produced an all sky astrometric and photometric
reference catalog (\refcat) to meet the needs of
the ATLAS project: all-sky uniformity, accuracy, and completeness for $6<m<19$ including the
galactic plane.

We have merged data from the three publicly released
star catalogs of Gaia, Pan-STARRS and
SkyMapper.  We supplemented this with new ATLAS
Pathfinder survey data, and we re-processed APASS images in the
southern sky to improve its photometric uniformity and
fidelity.  Addition of bright stars from Tycho-2 and the Yale Bright
Star catalog filled in the brightest stars that affect observations
even when saturated, and which are required by small
aperture, very wide-field instruments.  We have also added 2MASS
near-infrared $JHK$ magnitudes for each object, when they exist, to
remove the need for a user to do an extra catalog cross-match.

To our knowledge \refcat\
currently is the best single reference catalog for photometric
calibration of wide-field surveys. The
release of Gaia DR2 has obviously been a game changer for astrometry
and we recognize the exquisite calibration by including the most
important Gaia data in
the catalog table. The salient features of the catalog are
summmarised below; details on the content of \refcat\ are found in
Appendix~\ref{app:details}. 

\begin{itemize}
\item \refcat\ lists 991 million stars between
  $-1.5 < m \lesssim 19$ for which we provide $griz$ magnitudes.
  In this context $m$ means the brightest of $g$, $r$, and $i$ magnitude.
  We believe \refcat\ is virtually complete to $m<19$ over the entire sky.
  \refcat\ also repeats the Gaia DR2 coordinates,
  parallax, proper motion, $G$, $G_{BP}$, and $G_{RP}$ magnitudes, as
  well as Gaia estimates of photosphere temperature and extinction.  
  We augment this
  with \cite{SFD} extinction values, and we compute proximity statistics for
  each star's neighbors that may interfere with photometry.
  2MASS $JHK$ magnitudes are provided when available.  
\item All $griz$ photometry has been transformed to the Pan-STARRS $g_{P1}$,
  $r_{P1}$, $i_{P1}$, and $z_{P1}$ bandpasses.  Extension of the PS1
  photometry to catalogs south of Dec
  $-30$\deg\ and brighter than $m<14$ uses a two step procedure of
  identifying a small fraction of stars from the Gaia and 2MASS
  catalogs with well behaved photometric properties and determining
  relationships to convert Gaia and 2MASS photometry for these stars
  to the Pan-STARRS system (the GMP subset).
\item For reasons we do not fully understand, the GMP regressions
  produced residuals that depend on stellar sky density, possibly
  linked to a small photometry offset in Gaia when scanning high
  density fields. Whatever the reason, inclusion of a stellar density
  term in the regression equations removes the residual satisfactorily.
\item The GMP subset is then transformed as required to each catalog's
  bandpasses to provide zeropoints of observations or mean offsets on
  a square degree basis.  The uniformity of Gaia and 2MASS are the
  foundation for the uniformity of \refcat\ in the south and brighter
  than $m<14$.
\item The \refcat\ $griz$ magnitudes are weighted means from all
  contributing catalogues.  Pan-STARRS has the greatest signal to
  noise in the transformed $griz$ magnitudes, so north of Dec
  $-30$\deg\ and fainter than $m>14$ \refcat\ magnitudes are
  effectively Pan-STARRS DR1 magnitudes.
\item Comparisons of \refcat\ with the GMP subset indicate that the
  RMS systematic error in \refcat\ is on the order of 3~millimag, but excursions
  in sytematic error as large as 20~millimag do appear
  for selected square degrees near the galactic plane.
\item Comparison of \refcat\ with synthetic magnitudes from the STScI
  Calspec set of stars reveal no significant offsets in any of the
  bandpasses, implying that \refcat\ $griz$ agree with the Calspec
  system and Pan-STARRS bandpasses at the 5~millimag level or better.
\end{itemize}

The future operations of ATLAS in the south, ongoing SkyMapper
operations and data releases, and Gaia Data Release 3 (expected around
2021) will lead to improvements, and we envisage revising this data
product when these become available.  The Large Synoptic Survey
Telescope is expected to produce a definitive survey in the southern
hemisphere with its first data release due around 2023, and extremely
accurate photometry is part of its baseline requirement.  Although
LSST itself will saturate around $m<17$, a calibration telescope is
also planned which will be able to work to much brighter limits, so
these should provide a useful cross-check on both the bright and faint
end of \refcat\ when their calibrated results become available.

\acknowledgments

ATLAS observations and this work were supported by NASA grant NN12AR55G.
The AAVSO Photometric All-Sky Survey (APASS) was funded by the Robert
Martin Ayers Sciences Fund.
SJS acknowledges funding from STFC Grants  ST/P000312/1 and ST/N002520/1. 

This work has made use of data from the European Space Agency (ESA)
mission {\it Gaia} (\url{https://www.cosmos.esa.int/gaia}), processed
by the {\it Gaia} Data Processing and Analysis Consortium (DPAC,
\url{https://www.cosmos.esa.int/web/gaia/dpac/consortium}). Funding
for the DPAC has been provided by national institutions, in particular
the institutions participating in the {\it Gaia} Multilateral
Agreement.

Data products from the Two Micron All Sky Survey were used, which is a
joint project of the University of Massachusetts and the Infrared
Processing and Analysis Center/California Institute of Technology,
funded by the National Aeronautics and Space Administration and the
National Science Foundation.

The Pan-STARRS1 Surveys (PS1) and the PS1 public science archive have
been made possible through contributions by the Institute for
Astronomy, the University of Hawaii, the Pan-STARRS Project Office,
the Max-Planck Society and its participating institutes, the Max
Planck Institute for Astronomy, Heidelberg and the Max Planck
Institute for Extraterrestrial Physics, Garching, The Johns Hopkins
University, Durham University, the University of Edinburgh, the
Queen's University Belfast, the Harvard-Smithsonian Center for
Astrophysics, the Las Cumbres Observatory Global Telescope Network
Incorporated, the National Central University of Taiwan, the Space
Telescope Science Institute, the National Aeronautics and Space
Administration under Grant No. NNX08AR22G issued through the Planetary
Science Division of the NASA Science Mission Directorate, the National
Science Foundation Grant No. AST-1238877, the University of Maryland,
Eotvos Lorand University (ELTE), the Los Alamos National Laboratory,
and the Gordon and Betty Moore Foundation. 

The national facility capability for
SkyMapper has been funded through ARC LIEF grant LE130100104 from the
Australian Research Council, awarded to the University of Sydney, the
Australian National University, Swinburne University of Technology,
the University of Queensland, the University of Western Australia, the
University of Melbourne, Curtin University of Technology, Monash
University and the Australian Astronomical Observatory. SkyMapper is
owned and operated by The Australian National University's Research
School of Astronomy and Astrophysics. The survey data were processed
and provided by the SkyMapper Team at ANU. The SkyMapper node of the
All-Sky Virtual Observatory (ASVO) is hosted at the National
Computational Infrastructure (NCI). Development and support the
SkyMapper node of the ASVO has been funded in part by Astronomy
Australia Limited (AAL) and the Australian Government through the
Commonwealth's Education Investment Fund (EIF) and National
Collaborative Research Infrastructure Strategy (NCRIS), particularly
the National eResearch Collaboration Tools and Resources (NeCTAR) and
the Australian National Data Service Projects (ANDS).

\appendix

\section{\refcat\ Details}
\label{app:details}

\refcat\ is served from 
MAST at the Space Telescope Science Institute 
(URL {\url{https://archive.stsci.edu/prepds/atlas-refcat2/}).
All our data products are available at MAST via
\dataset[doi:10.17909/t9-2p3r-7651]{http://doi.org/10.17909/t9-2p3r-7651}.
STScI provides \refcat\ in a variety of formats: the basic bzip2 compressed
tar archives with integer fields described here, other types of file compression, 
and a version
that replaces the scaled integers with values that have the correct units.
STScI also provides an interface into a database that permits cone searches
and other queries.  Finally STScI will match \refcat\ against the
UV photometry from the Galex satellite and IR photometry from the WISE
satellite, and provide those data as well.

\refcat\ is normally provided as a set of bzip2 compressed tar archives of
64800 files, one for each (coordinate) square degree in the sky.  The file names reflect
the coordinate location, {\tt rrr+dd.rc2}.  For example {\tt 270-20.rc2} is the
square degree with $270\le RA < 271$ and $-20\le Dec < -19$.  The stars in
each square degree file are sorted by increasing RA.  The data are given in
comma separated variable format (CSV), using scaled integers for all real numbers
as described in Table~\ref{tab:rccols}. Abbreviations include ``10ndeg'' for $10^{-8}$~degree, ``10uas'' for $10^{-5}$~arcsecond, ``mas'' for $10^{-3}$~arcsecond, and ``mmag'' for $10^{-3}$ magnitude.

\begin{table}[htp]
\caption{\refcat\ table columns}
\begin{center}
\begin{tabular}{rlrlrl}
\hline
Field & Name   & Entry         & Units        & Meaning        & Description \\
\hline
 1 & {\tt RA}     &  28000001672  & {\tt [10ndeg]}     &  280.00001672\deg  & RA from Gaia DR2, J2000, epoch 2015.5 \\
 2 & {\tt Dec}    & $-$1967818581 & {\tt [10ndeg]}     & $-19.67818581$\deg & Dec from Gaia DR2, J2000, epoch 2015.5 \\
 3 & {\tt plx}    &    98         & {\tt [10uas]}      &  0.98~mas      & Parallax from Gaia DR2 \\
 4 & {\tt dplx}   &    10         & {\tt [10uas]}      & 0.10~mas       & Parallax uncertainty \\
 5 & {\tt pmra}   &   114         & {\tt [10uas/yr]}   &  1.14~mas/yr   & Proper motion in RA from Gaia DR2 \\
 6 & {\tt dpmra}  &    16         & {\tt [10uas/yr]}   & 0.16~mas/yr    & Proper motion uncertainty in RA \\
 7 & {\tt pmdec}  & -1460         & {\tt [10uas/yr]}   & -14.60~mas/yr  & Proper motion in Dec from Gaia DR2 \\
 8 & {\tt dpmdec} &    15         & {\tt [10uas/yr]}   &  0.15~mas/yr   & Proper motion uncertainty in Dec \\
 9 & {\tt Gaia}   & 15884         & {\tt [mmag]}       & 15.884         & Gaia DR2 $G$ magnitude \\
10 & {\tt dGaia}  &     1         & {\tt [mmag]}       &  0.001         & Gaia DR2 $G$ magnitude uncertainty \\
11 & {\tt BP}     & 16472         & {\tt [mmag]}       & 16.472         & Gaia $G_{BP}$ magnitude \\
12 & {\tt dBP}    &    10         & {\tt [mmag]}       &  0.010         & Gaia $G_{BP}$ magnitude uncertainty \\
13 & {\tt RP}     & 15137         & {\tt [mmag]}       & 15.137         & Gaia $G_{RP}$ magnitude \\
14 & {\tt dRP}    &     1         & {\tt [mmag]}       &  0.001         & Gaia $G_{RP}$ magnitude uncertainty \\
15 & {\tt Teff}   &  4729         & {\tt [K]}          & 4729~K         & Gaia stellar effective temperature \\
16 & {\tt AGaia}  &   895         & {\tt [mmag]}       & 0.895          & Gaia estimate of $G$-band extinction for this star\\
17 & {\tt dupvar} &     2         &                    & 2              &
 \begin{minipage}[t]{0.4\columnwidth} %
 Gaia flags coded as {\tt CONSTANT} (0), {\tt VARIABLE} (1), or {\tt NOT\_AVAILABLE} (2) {\tt + 4*DUPLICATE}
  \end{minipage} \\
18 & {\tt Ag}     &  1234         & {\tt [mmag]}       & 1.234          & SFD estimate of total column $g$-band extinction \\
19 & {\tt rp1}    &    50         & {\tt [0.1\arcsec]} &  5.0\arcsec    & Radius where cumulative $G$ flux exceeds 0.1$\times$ this star \\
20 & {\tt r1}     &    50         & {\tt [0.1\arcsec]} &  5.0\arcsec    & Radius where cumulative $G$ flux exceeds 1$\times$ this star \\
21 & {\tt r10}    &   155         & {\tt [0.1\arcsec]} & 15.5\arcsec    & Radius where cumulative $G$ flux exceeds 10$\times$ this star \\
22 & {\tt g}      & 16657         & {\tt [mmag]}       & 16.657         & Pan-STARRS $g_{P1}$ magnitude \\
23 & {\tt dg}     &    10         & {\tt [mmag]}       &  0.010         & Pan-STARRS $g_{P1}$ magnitude uncertainty \\
24 & {\tt gchi}   &    23         & {\tt [0.01]}       &  0.23          & $\chi^2/DOF$ for contributors to $g$\\
25 & {\tt gcontrib}  &    1f         & {\tt [\%02x]}      & 00011111       & Bitmap of contributing catalogs to $g$\\
26 & {\tt r}      & 15915         & {\tt [mmag]}       & 15.915         & Pan-STARRS $r_{P1}$ magnitude \\
27 & {\tt dr}     &    12         & {\tt [mmag]}       &  0.012         & Pan-STARRS $r_{P1}$ magnitude uncertainty \\
28 & {\tt rchi}   &    41         & {\tt [0.01]}       &  0.41          & $\chi^2/DOF$ for contributors to $r$\\
29 & {\tt rcontrib}  &    3f         & {\tt [\%02x]}      & 00111111       & Bitmap of contributing catalogs to $r$ \\
30 & {\tt i}      & 15578         & {\tt [mmag]}       & 15.578         & Pan-STARRS $i_{P1}$ magnitude \\
31 & {\tt di}     &    10         & {\tt [mmag]}       &  0.010         & Pan-STARRS $i_{P1}$ magnitude uncertainty \\
32 & {\tt ichi}   &    49         & {\tt [0.01]}       &  0.49          & $\chi^2/DOF$ for contributors to $i$\\
33 & {\tt icontrib}  &    0f         & {\tt [\%02x]}      & 00001111       & Bitmap of contributing catalogs to $i$ \\
34 & {\tt z}      & 15346         & {\tt [mmag]}       & 15.346         & Pan-STARRS $z_{P1}$ magnitude \\
35 & {\tt dz}     &    12         & {\tt [mmag]}       &  0.012         & Pan-STARRS $z_{P1}$ magnitude uncertainty \\
36 & {\tt zchi}   &     0         & {\tt [0.01]}       &  0.00          & $\chi^2/DOF$ for contributors to $z$\\
37 & {\tt zcontrib}  &    06         & {\tt [\%02x]}      & 00000110       & Bitmap of contributing catalogs to $z$ \\
38 & {\tt nstat}  &     0         &                    & 0              & Count of $griz$ deweighted outliers \\
39 & {\tt J}      & 14105         & {\tt [mmag]}       & 14.105         & 2MASS J magnitude \\
40 & {\tt dJ}     &    36         & {\tt [mmag]}       & 0.036          & 2MASS J magnitude uncertainty \\
41 & {\tt H}      & 14105         & {\tt [mmag]}       & 14.105         & 2MASS H magnitude \\
42 & {\tt dH}     &    53         & {\tt [mmag]}       & 0.053          & 2MASS H magnitude uncertainty \\
43 & {\tt K}      & 13667         & {\tt [mmag]}       & 13.667         & 2MASS K magnitude \\
44 & {\tt dK}     &    44         & {\tt [mmag]}       & 0.044          & 2MASS K magnitude uncertainty \\
\hline
\end{tabular}
\end{center}
\label{tab:rccols}
\end{table}%

When a magnitude is not available (for
example 2MASS at the faint end or $G$ for non-Gaia stars) the magnitude and its uncertainty are set to 0,
otherwise the magnitude uncertainty is given as at least 1~millimag.

The $g$ band total column extinction 
{\tt Ag} is computed from the $E(B{-}V)$ values of \cite{SFD}, multiplied by
0.88 as recommended by \cite{Schlafly11}, and also by
$ Ag/E(B{-}V) = 3.613 - 0.0972 (g{-}i) + 0.0100 (g{-}i)^2$
\citep{psphot}.

The proximity statistics {\tt rp1},  {\tt r1}, and {\tt r10} are derived by summing
the cumulative $G$ band flux of all Gaia stars as a function of
distance from each star, and reporting
the radius where this flux first exceeds 0.1 ({\tt rp1}), 1 ({\tt r1}), and 10 
({\tt r10}) times the flux of the star.  These are given the value 999
(99.9\arcsec) when a star is so isolated that the cumulative flux never reaches
the threshold within the 36\arcsec\ search radius.

The {\tt griz}-{\tt contrib} entries identify contributors to the $griz$ magnitudes.  Bits 0--7
are set when a catalog contributes to the statistical average 
with magnitude uncertainty less than 0.2: Gaia DR2 (bit 0), GMP (bit 1), Pan-STARRS (bit 2),
SkyMapper (bit 3), Pathfinder (bit 4), APASS (bit 5), APASS DR9 (bit 6), and Tycho-2/BSC (bit 7).
For example the code {\tt rcontrib}=06 implies that $r$ contributions
with uncertainty less than 0.2 mag came from GMP and Pan-STARRS.

Gaia DR2 does not include all bright stars; Polaris is missing, for
example, as well as a $m{\sim}12$ star at RA 93.7759, Dec +15.0451.
Stars found in the contributing catalogs
that are no closer than 3.6\arcsec\ (PanSTARRS, SkyMapper,
and Tycho/BSC) or 10.8\arcsec\ (Pathfinder and APASS) to the nearest Gaia star are added to
\refcat\ with zero values for all Gaia-specific quantities.  A real star
can therefore be inhibited from inclusion because Gaia DR2 lists a
faint star nearby, except for Tycho/BSC for which the Gaia match must be within 2
magnitudes of the non-Gaia candidate.  A non-Gaia
star may be identified in \refcat\ because it will always have {\tt dGaia = 0}.

Examining a statistical sample of these non-Gaia objects, we see a
change in behavior around $g\sim12$, where the Tycho-2 catalog ends.
About 78\% of the non-Gaia inclusions in \refcat\ brighter than $g\sim12$ 
really are stars, about 18\% are double
star blends mostly from APASS DR9, and the remainder are bright
galaxies, completely false triggers (often in the outskirts of a bright
star), or coordinates that did not link with the Gaia stars (possibly a transient).  The 
double star blends almost always have the individual stars also present 
in \refcat\ as Gaia-matching entries.  About 0.25\% of the stars with $m<10$
in \refcat\ do not appear in Gaia DR2; 8946 stars in \refcat\ 
brighter than $m<12.5$ are missing from Gaia DR2.

Fainter than $g\sim12$ the fractions of stars, galaxies, and false triggers
near bright stars change discontinuously.  About 31\% appear to be real stars, 
27\% appear to be diffraction
artifacts from bright stars, 23\% are galaxies, and 19\% are probably transients, 
eruptive stars, or errors.  About 0.1\% of \refcat\ is not found in Gaia DR2
for these fainter sources.

A \refcat\ user should therefore use these non-Gaia objects according to
application.  If a relatively complete sample of $m>12$ stars is needed for
an astrometric or photometric solution there is no reason to include
any non-Gaia stars and selecting on {\tt dGaia > 0} may be advisable.  If the
user wants to know whether there is a star with $m<3$ within a degree, then
using the non-Gaia entries is mandatory.  If a detection appears
in a difference image it is a good idea to check the non-Gaia entries
as well for the identity of the object, but there is a chance ($\sim$0.1\%) that a
non-Gaia entry may not really be a bona fide star at the specified location.

A small fraction of Refcat2 entries are galaxies, inherited from Gaia DR2
and Pan-STARRS.  Comparing a subset of Refcat2 objects with SDSS DR12 star/galaxy
identifications and using Pan-STARRS (Kron-PSF) magnitudes, we find a galaxy
contamination rate that depends on magnitude and galactic latitude.  For $b^{II}>60\deg$
and $m<17$ the contamination is about 1.5\%, rising to $\sim$10\% for $m<19$.
For $b^{II}<30\deg$ the contamination is about 2\% for $m<19$.  Overall for
$m<16$ the galaxy fraction is less than 1\%.  Virtually all galaxies can be
rejected by selecting objects for which 
Gaia provides a non-zero proper motion uncertainty, {\tt dpmra} and {\tt dpmdec},
at the cost of about 0.7\% of all real stars.

Parenthetically, users interested in transforming Refcat2 to SDSS should {\it not}
use the relations of \cite{psphot} which are appropriate for SDSS DR7.  SDSS 
photometry evolves with each release, so although the color terms of \cite{psphot}
may still be correct, the offsets are not.

The catalog is distributed in five magnitude chunks without overlap, 
$gri<16$ ({\tt 00\_m\_16.tbz}, 105M stars, 5.9GB), 
$16\le gri<17$ ({\tt 16\_m\_17.tbz}, 107M stars, 5.6GB), 
$17\le gri<18$ ({\tt 17\_m\_18.tbz}, 204M stars, 9.8GB), 
$18\le gri<19$ ({\tt 18\_m\_19.tbz}, 369M stars, 17GB), and 
$19\le gri$ ({\tt 19\_m\_20.tbz}, 206M stars, 8.7GB).  Any star that has 
{\tt ($16\le m<17$)} where $m$ is the brightest of $g$, $r$, and $i$
lies in the second chunk, etc.  There are red stars with $g\ge16$ that appear in
the first chunk because $i<16$, but a user who wants a complete
sample with $g<17$, for example, needs only examine the first two chunks.
Stars appear in the fifth chunk with $gri\ge19$ because our Pan-STARRS
selection also includes red stars with $z<19$, but that chunk is very
incomplete.

The entire compressed catalog amounts to about 50 bytes per star (about
1 byte per field),
depending on how many fields are populated. The expectation is that 
users will select a subset of stars and fields chosen according to 
their application, which can greatly reduce the size and increase the 
access speed of a database table.  The data are partitioned into square 
degree files so that a survey such as ATLAS, whose field of view will typically 
touch $\sim$40 such files, can dispense with a database and simply use 
the file system to rapidly return all the stars in a given exposure.

\section{Pan-STARRS catalog query}
\label{app:ps1}

This is the query used to select objects with from Pan-STARRS DR1 at the STScI MAST 
archive\footnote{\tt http://mastweb.stsci.edu/ps1casjobs/}, 
as described in Section\,\ref{sec:ps1}. 
~\\
\noindent
{\tt select raMean, decMean, gMeanPSFMag, gMeanPSFMagErr, rMeanPSFMag, rMeanPSFMagErr, \\ 
  iMeanPSFMag, iMeanPSFMagErr, zMeanPSFMag, zMeanPSFMagErr, yMeanPSFMag, yMeanPSFMagErr \\
  from ObjectThin join MeanObject on objectThin.uniquePspsOBid = meanObject.uniquePspsOBid \\
  join stackObjectThin on objectThin.objID = stackObjectThin.objID \\
  where bestdetection = 1 and primarydetection = 1 and ((gKronMag $<$
  19 and gKronMag $>$ 0) \\
  or (rKronMag $<$ 19 and rKronMag $>$ 0 ) or (iKronMag $<$ 19 and iKronMag $>$ 0) \\
  or (zKronMag $<$ 19 and zKronMag $>$ 0)) }

This query does not attempt to do any star-galaxy separation. The Pan-STARRS 
catalog flag \texttt{QF\_OBJ\_EXT} (which implies a likely extended object) indicates that 
about 12\% of sources could potentially be galaxies.  However only a tiny fraction of
these sources do not match Gaia ($\sim0.1$\%), and only a quarter of those appear to
be galaxies, so the galaxy contamination should be minimal.

\clearpage
% default system hardware and software papers 

\end{document}